\documentclass[a4paper,fleqn,usenatbib]{mnras}

\usepackage{newtxtext,newtxmath}

\usepackage[T1]{fontenc}
\usepackage{ae,aecompl}

\usepackage{graphicx}	
\usepackage{amsmath}	
\usepackage{amssymb}	
\usepackage[singlelinecheck=off,justification=raggedright,position=top]{subfig}
\usepackage{hyperref}

\title[MORGOTH: SFH estimation using the HB]{MORGOTH: incorporating horizontal branch modelling into star formation history determinations}

\author[A. Savino et al.]{
A. Savino,$^{1,2}$\thanks{E-mail: A.savino@rug.nl}
T.J.L. de Boer,$^{3}$
M. Salaris$^{1}$
and E. Tolstoy$^{2}$
\\
$^{1}$Astrophysics Research Institute, Liverpool John Moores University, IC2, Liverpool Science Park, 146 Brownlow Hill, L3 5RF Liverpool, UK\\
$^{2}$Kapteyn Astronomical Institute, University of Groningen, Postbus 800, 9700 AV Groningen, The Netherlands\\
$^{3}$Department of Physics, University of Surrey, Guildford, GU2 7XH, UK
}

\date{Accepted XXX. Received YYY; in original form ZZZ}

\pubyear{2016}

\begin{document}
\label{firstpage}
\pagerange{\pageref{firstpage}--\pageref{lastpage}}
\maketitle

\begin{abstract}
We present a new method that incorporates the horizontal branch morphology into synthetic colour-magnitude diagram based star formation history determinations. This method, we call {\small MORGOTH}, self-consistently takes into account all the stellar evolution phases up to the early asymptothic giant branch, flexibly modelling red giant branch mass loss. We test {\small MORGOTH} on a range of synthetic populations, and find that the inclusion of the horizontal branch significantly increases the precision of the resulting star formation histories. When the main sequence turn-off is detected, {\small MORGOTH} can fit the star formation history and the red giant branch mass loss at the same time, efficiently breaking this degeneracy. As part of testing {\small MORGOTH}, we also model the observed colour-magnitude diagram of the well studied Sculptor dwarf spheroidal galaxy. We recover a new more detailed star formation history for this galaxy. Both the new star formation history and the red giant branch mass loss we determined for Sculptor with {\small MORGOTH} are in good agreement with previous analyses, thus demonstrating the power of this new approach.
\end{abstract}

\begin{keywords}
galaxies: star formation -- galaxies: evolution -- galaxies: stellar content -- Hertzsprung-Russell and C-M diagrams -- stars: horizontal-branch -- stars: mass-loss
\end{keywords}



\section{Introduction}

Determining detailed star formation histories (defined as the star formation rate as a function of age and metallicity -- SFHs)  for a variety of different types of galaxies is important to the understanding of galaxy formation and evolution. The SFH of galaxies in the local Universe, going back to the earliest times, allows a  comparison with predictions, over the same time frame, from cosmological and galaxy evolution models \citep[see, e.g.,][ for models of Local Volume galaxy properties]{Lanfranchi04,Salvadori08,Romano13,Starkenburg13,Garrison-kimmel14,Fattahi16,Sorce16}. In this way we can constrain the conditions in the local Universe when the Milky Way and its satellites were forming.

Resolving individual stars, in a stellar population, down to the main sequence turn-off (MSTO), allows a detailed and accurate SFH to be determined. Unfortunately, with current observational capabilities we are only able to resolve a small sample of galaxies, primarily dwarf galaxies in the Local Group \citep[e.g.,][]{Tolstoy09, Weisz11}. Even so, the information that we can extract from these few nearby dwarf galaxies is very valuable and complements what we can learn from the study of unresolved, more distant galaxies \citep[e.g.,][]{Gallazzi08,Boylan-Kolchin16,Goddard17}. Moving beyond the satellites of the Milky Way, detecting the main sequence of the old population becomes challenging, and photometric data is often only able to reach down to the upper red giant branch (RGB) and the horizontal branch (HB) of galaxies \citep[e.g.,][]{Martin16,Martin17}.

To determine the SFH of a resolved galaxy, a common approach is to build synthetic colour-magnitude diagrams (CMDs) to compare observed and predicted stellar distributions \citep[e.g.,][]{tosi91,tolstoy96,Aparicio97,Dolphin97,Harris01,Cignoni10}. An important contribution towards quantitative and reliable SFH determinations came with the work of \citet{Dolphin02}, who formalized many of the numerical challenges in SFH recovery, and performed detailed modelling of an homogeneous Hubble Space Telescope (HST) dataset of nearby galaxies. This framework was further developed with a thorough characterization of the typical uncertainties in SFH determinations by \citet{Dolphin12} and \citet{Dolphin13}. The SFH determinations for nearby galaxies, using synthetic techniques, are extensive, thanks especially to the exquisite sensitivity and resolving power provided by the advent of HST \citep[e.g.,][]{Aloisi99,Cole07,Monelli10b,Monelli10a,Weisz11,Weisz14a,Sacchi16}.

One of the limitations of current SFH determinations of old stellar populations in the Local Group is that synthetic diagrams are typically only generated up to the tip of the RGB. Therefore, they neglect the later stages of stellar evolution, most importantly the HB phase. The main reason for this has been that RGB mass loss plays a strong role in shaping the morphology of the HB. As the mass loss phenomenon has proven difficult to characterize, this dependence has made predicting the HB of a given stellar population challenging \citep[e.g.,][and references therein]{Gratton10}.

Accurately modelling the HB, taking into account the uncertain value of RGB mass loss, is a promising way forward to improve the SFH for nearby and distant galaxies. HB stars are much brighter than their MSTO counterparts, making them less affected by photometric uncertainties. This means that the HB morphology can be accurately characterized out to the edge of the Local Group and beyond. Theoretical calculation shows that the photometric properties of HB stars depend mostly on the metallicity and the stellar mass \citep[e.g.,][]{Iben70}. For a fixed chemical composition, the mass of HB stars is determined by the age of the stellar population and the amount of mass lost during the previous RGB phase. Thus, at fixed mass loss the HB morphology depends only on the SFH of the system. Furthermore, when the MSTO morphology is included, then the modelling of the HB has the potential to dramatically mitigate the age-metallicity degeneracy, as two stellar populations with different age-metallicity combination and the same MSTO luminosity will look quite different on the HB. For these reasons, the information contained in the HB can be used to significantly improve the recovered SFH, provided that we can accurately model the mass loss \citep{savino15}.

Previous studies have used the HB to constrain SFHs both in the Milky Way \citep{Preston91,Santucci15} and in external galaxies \citep{Schulte-ladbeck02,Rejkuba11,Grocholski12}. In particular, the analysis of the HB has been crucial to constrain the stellar population properties of galaxies beyond our nearest Galactic companions. However, determining an accurate star formation rate, comparable to measurements from the MSTO,  from the HB has been difficult to achieve. This is because these previous analyses necessarily employed theoretical isochrones for the HB modelling. These are built assuming a specific, only loosely constrained, mass loss efficiency. Not allowing this parameter to vary has a big impact on the measured SFH from the HB.

In this paper we present a new SFH determination technique that flexibly and consistently takes into account the HB morphology when analysing the CMD of a galaxy, allowing a variation in the RGB mass loss. This approach provides precise SFHs and accurate measurements of the RGB mass loss. The starting point of {\small MORGOTH} (\textbf{M}odelling \textbf{O}f \textbf{R}esolved \textbf{G}alaxies with \textbf{O}ptimized \textbf{T}urn-off and \textbf{H}B synthesis), was the routine {\small TALOS} \citep{deBoer12}, which has been modified to include synthetic HBs, to handle the effect of mass loss and to recover a more detailed SFH, using the information of all the evolutionary phases up to the end of the early asymptotic giant branch. 

In \S\ref{talos} we give a brief description of how {\small TALOS} works, in \S\ref{morgoth} we describe {\small MORGOTH}, in \S\ref{test} we present several performance tests using mock observations and in \S\ref{Sculptor} we model the stellar population of the Sculptor dwarf spheroidal galaxy (dSph). In \S\ref{conclusion} we summarize our work.

\section{Talos, modelling the MSTO region}
\label{talos}
{\small TALOS} is a SFH determination routine presented and verified in \citet{deBoer12}. The algorithm of this code has been developed according to the prescriptions laid out in \citet{Dolphin02}. For a complete description, see \citet{deBoer12}. Here we briefly describe the main features.

{\small TALOS} determines a galaxy SFH by comparing an observed CMD to a set of synthetic simple stellar population models, which are generated on a fine grid of age and metallicity (in this paper the term metallicity always refers to [Fe/H]). Parameters like the binary fraction, the [$\alpha$/Fe]~vs~[Fe/H] relation and the initial mass function (IMF) are taken into account. If available, spectroscopic metallicites can also be added as an additional constraint. The synthetic metallicity distribution functions (MDFs) are generated, sampling the model CMDs as the spectroscopic measurements. In order to accurately compare models and observations, observational effects are added to the synthetic CMDs to match the observations. Such effects include distance, reddening, photometric uncertainties and completeness fractions across the CMD. Measurement errors and completeness levels are determined from artificial star tests. 

After transforming the CMDs into Hess diagrams (defined as the density map of stars across the CMD, given a specific CMD binning scheme), {\small TALOS} searches for the linear combination of models that minimize the Poisson analogue of the unreduced $\chi^2$ function:
\begin{equation}
\chi^2_{P}=2\sum_{i}{(m_{i}-n_{i}+n_i\ln{\frac{n_{i}}{m_{i}}})}
\end{equation}

Where $m_{i}$ is the number of stars in a synthetic Hess bin, and $n_{i}$ is the observed star count in the same bin. The spectroscopic MDF comparison can be included in the $\chi^2_P$ evaluation. The MDFs are rescaled, to compensate for the CMD having many more stars and consequently dominating the final $\chi^2_P$. Given the regularity of the $\chi^2_P$ surface built in this way, the minimum can be located through the Fletcher-Reeves-Polak-Ribiere ({\small FRPR}) algorithm \citep{Press92}, which makes use of a conjugate gradient technique. This method is very fast, as it makes use of the function derivatives to find the minimum.

The uncertainties are evaluated by fitting the SFH for a range of different CMD sampling and parameter space sampling combinations (the former relates to the CMD binning choice while the latter relates to the binning of the age-metallicity grid). For each combination of CMD and parameter space sampling, a solution is evaluated and is used to estimate the uncertainties related to data sampling. These arise because the observed stellar population is finite in size and is just a random realization of the underlying statistical population. In this way, the final SFH solution is the average of at least 100 different solutions, from which the standard deviation is used to estimate the uncertainties on the star formation rates.

\section{MORGOTH, including the horizontal branch}
\label{morgoth}

The main complication of including the HB morphology (when modelling the SFH of a galaxy) is that the amount of mass lost by RGB stars during their evolution is uncertain.Thus, in principle, exploring the effects of different mass loss assumptions requires to generate a complete synthetic CMD for each mass loss prescription, as done, for instance, in \citet{Salaris13}. Such an approach has a high computational cost and makes an extensive exploration of the mass loss parameter space challenging.

Fortunately, another approach is possible, because the helium core mass in low mass stars ($M \lesssim 1.5 M_{\odot}$) at the start of the helium burning phase is very weakly dependent on the progenitor mass \citep[see, e.g.,][]{sc}. In addition, the progenitor RGB evolution is essentially insensitive to mass loss, unless extremely high values are considered. These properties imply that a theoretical HB evolutionary track, of a given stellar mass and chemical composition, is independent of the combination of progenitor mass (i.e., age of the stellar population) and total RGB mass loss. This property of HB stars is commonly exploited in synthetic stellar population generations \citep[e.g., ][]{Salaris13,savino15,Tailo16}.

Using this technique in {\small MORGOTH}, we treat the HB modelling in a straightforward way. We compute a grid of synthetic simple stellar populations, using the same procedure of {\small TALOS}, which are generated up to the tip of the RGB. This grid covers a range of stellar population ages and metallicities. These simple stellar population models are then complemented by a grid of synthetic HB models, which are calculated for different metallicities and HB stellar masses (see \S~\ref{modcomp} for details). These HB models contain stars between the onset of quiescent helium burning and the first thermal pulse on the asymptotic giant branch. For any arbitrary mass loss prescription, we can map the HB grid onto the simple stellar population grid. After appropriately rescaling the stellar density, the two synthetic models are combined to create a self-consistent Hess diagram that includes the helium burning phase. In this way we greatly reduce the number of the required synthetic CMD computations.

\begin{figure}
	\includegraphics[width=\columnwidth]{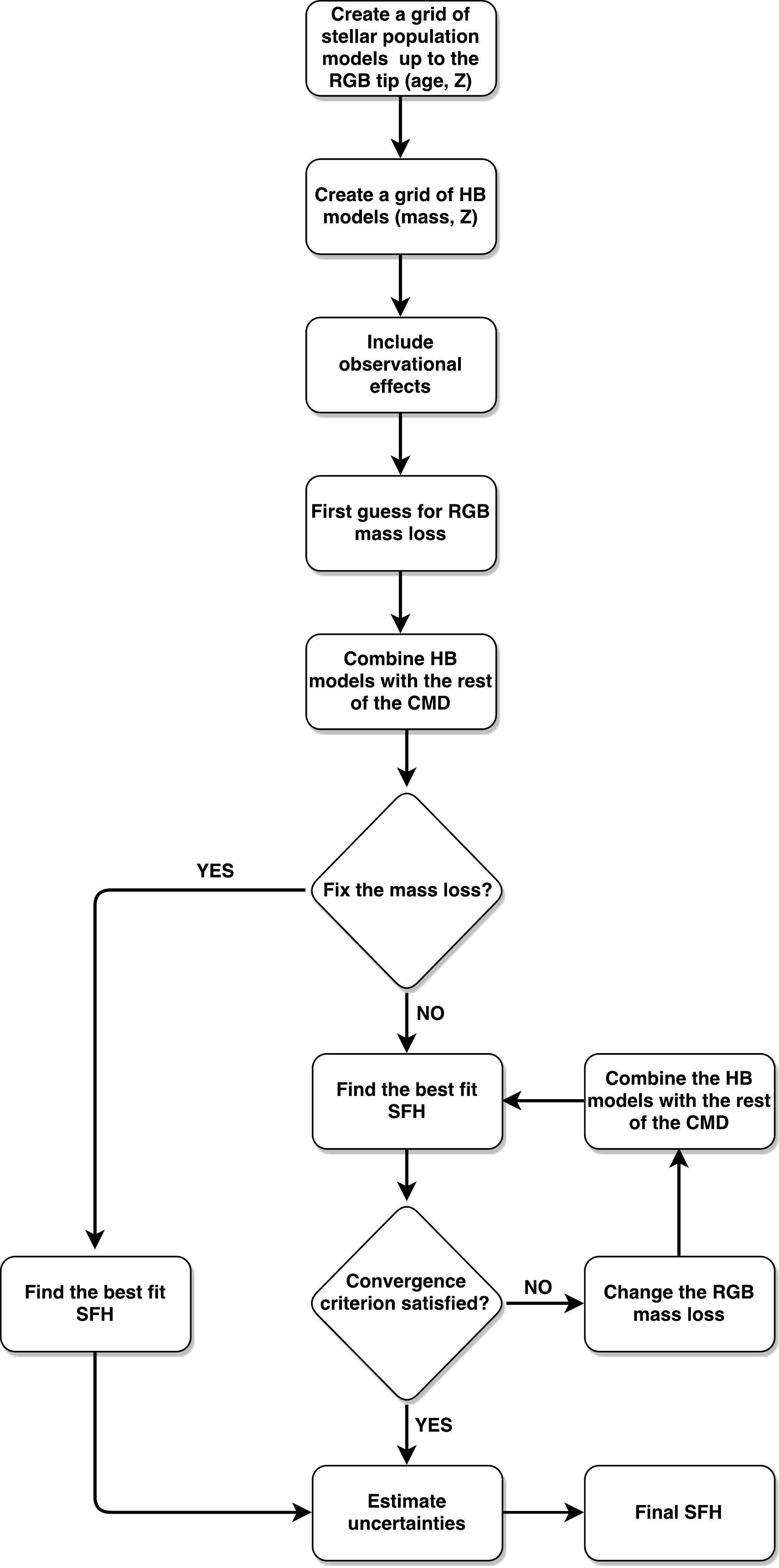}
    \caption{Flow chart representing the main procedural steps of {\small MORGOTH}.}
    \label{fig:flowchart}
\end{figure}

 In the rest of this section we will provide a general description of the {\small MORGOTH} approach, which is also schematized in Figure \ref{fig:flowchart}.

\subsection{HB model computation}
\label{modcomp}
The stellar evolution library we use to compute all the synthetic models is the BaSTI stellar library \citep{basti,Pietrinferni06}. These models have been successfully tested and used to reproduce the HB properties of globular clusters and resolved galaxies \citep[e.g.,][]{Cassisi03,Dalessandro11,Dalessandro13,Salaris13}. To increase our method's versatility, we extended the HB track grid, computing additional tracks, reaching a mass of $1.5 M_{\odot}$ at the beginning of the He-burning phase (A. Pietrinferni, private communication). With this extended grid we can create synthetic CMDs of intermediate age stellar populations (as young as $\sim 2 \,Gyr$), that typically present a red clump instead of an HB. In the following we describe how we calculate our grid of synthetic HB models.

For any given simple stellar population, we assume that RGB stars lose a total amount of mass $\Delta M$, which has a Gaussian distribution with standard deviation $\sigma_{\Delta M}$. The extent of this Gaussian spread is a chosen parameter. Recent studies of globular clusters and dwarf galaxies suggest that the dispersion is of the order of a few thousandths of solar mass, at fixed metallicity \citep{Caloi08, Salaris13, Tailo16}. In all the synthetic CMDs shown in this paper, we keep this value fixed at $\sigma_{\Delta M} = 0.005 M_{\odot}$. Thus the resultant HB stars will have a Gaussian distribution of mass on the zero age horizontal branch (ZAHB), with an average value $\overline{M}_{\scriptscriptstyle{ZAHB}}$. We generate a set of synthetic HB models, covering a grid of [Fe/H] and $\overline{M}_{\scriptscriptstyle{ZAHB}}$ values. 

\begin{figure}[t!]
	\includegraphics[width=\columnwidth]{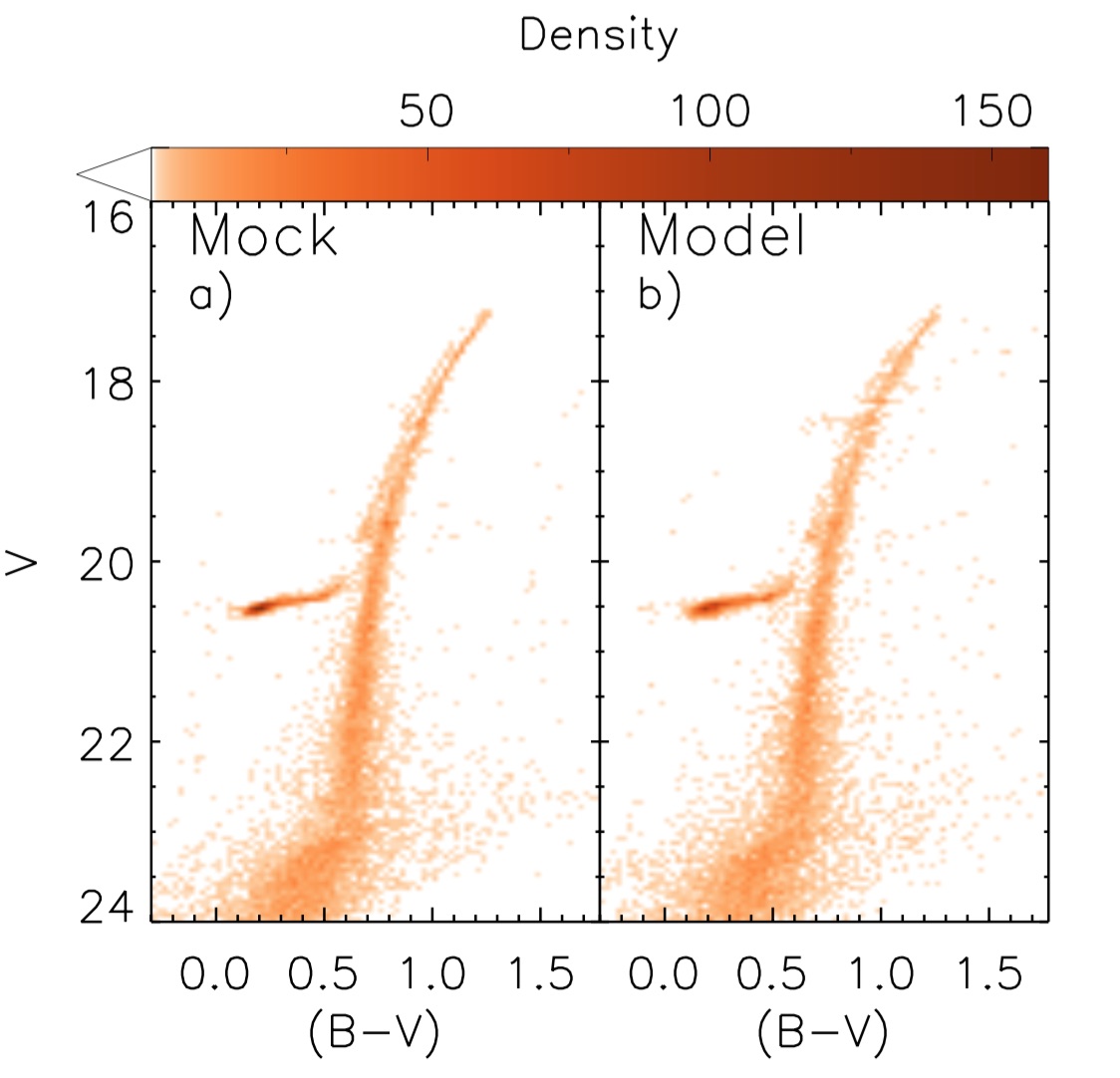}
    \caption{\textit{a)}: the (B - V) vs V Hess diagram of our old and metal poor mock SSP (t = 12.5 Gyr, [Fe/H] = -2.5 dex). \textit{b)}: The best fit Hess diagram recovered with {\small MORGOTH}.}
    \label{fig:CMDomp}
\end{figure}

To populate a synthetic HB model, we need to calculate a distribution of stellar masses and a distribution of ages since the start of the helium burning. Aside from the obvious constraint that the average mass on the ZAHB should be equal to $\overline{M}_{\scriptscriptstyle{ZAHB}}$, the synthetic HB model has to satisfy two more conditions. The first one is that stars in later phases of the helium burning should be, on average, more massive than stars on earlier phases. This is because, at fixed mass loss, stars near the end of the helium burning come from more massive progenitors. The second constraint is that, when the synthetic HB is matched to a simple stellar population CMD, the number of stars on the HB and the stellar mass distribution have to be consistent with the adopted IMF.

These constraints are satisfied by adopting the following procedure. For each value of [Fe/H] and $\overline{M}_{\scriptscriptstyle{ZAHB}}$ we assume a realistic mass loss value \citep[we adopt the prescription from][ but the choice is not particularly critical]{Origlia14} to calculate the progenitor mass that produces $\overline{M}_{\scriptscriptstyle{ZAHB}}$ . We call this value $M^{\scriptscriptstyle{prog}}_{\scriptscriptstyle{ZAHB}}$. We then consider an age range $\Delta t=200 Myr$ (the exact value is not important, as long as it is larger than the maximum HB lifetime) and determine the range of initial masses whose age at the tip of the RGB is in the range between the age of $M^{\scriptscriptstyle{prog}}_{\scriptscriptstyle{ZAHB}}$ denoted as $t_0$, and $t_0-\Delta t$. We extract a mass distribution for this range of progenitors according to the chosen IMF, and assign to them the corresponding HB mass by adding the selected value of the RGB mass loss (with a Gaussian distribution). For each HB mass we calculate its location along the HB phase considering that its age from the ZAHB is equal to the age difference at the tip of the RGB between its progenitor and the progenitor of $\overline{M}_{\scriptscriptstyle{ZAHB}}$. HB masses whose age would be larger than their HB lifetime are considered to be evolved off the HB.

 When the mass loss is changed, we calculate a correction factor, for each HB model, that takes into account the change in the progenitor masses. This factor is the ratio, coming from the IMF, of the new progenitor number abundance to the original progenitor number abundance. We use this factor to scale the HB density distribution so that our models are always consistent with the adopted IMF. However, due to the short lifetime of HB stars, this correction is very small.

\subsection{HB matching to the rest of the CMD}

Given the two grids of models we generate (one for the HBs and one for the rest of the CMDs), we are able create synthetic CMDs that include the HB and the early asymptotic giant branch. Our method is able to create synthetic CMDs for any arbitrary mass loss prescription, as the only requirement is to calculate $\overline{M}_{\scriptscriptstyle{ZAHB}}$ for a given population, to chose the appropriate HB model and to rescale its Hess diagram consistently with the IMF. The simple stellar population model and its corresponding HB are then merged into a single Hess diagram. As the HB grid is discrete, there might be small differences between the $\overline{M}_{\scriptscriptstyle{ZAHB}}$ of a given population and the closest HB model. For this reason, a very fine mass sampling of the HB grid is preferred. We find a compromise between mass sampling and computation speed with a mass sampling step of $0.005 M_{\odot}$

\begin{figure}[t!]
	\includegraphics[width=\columnwidth]{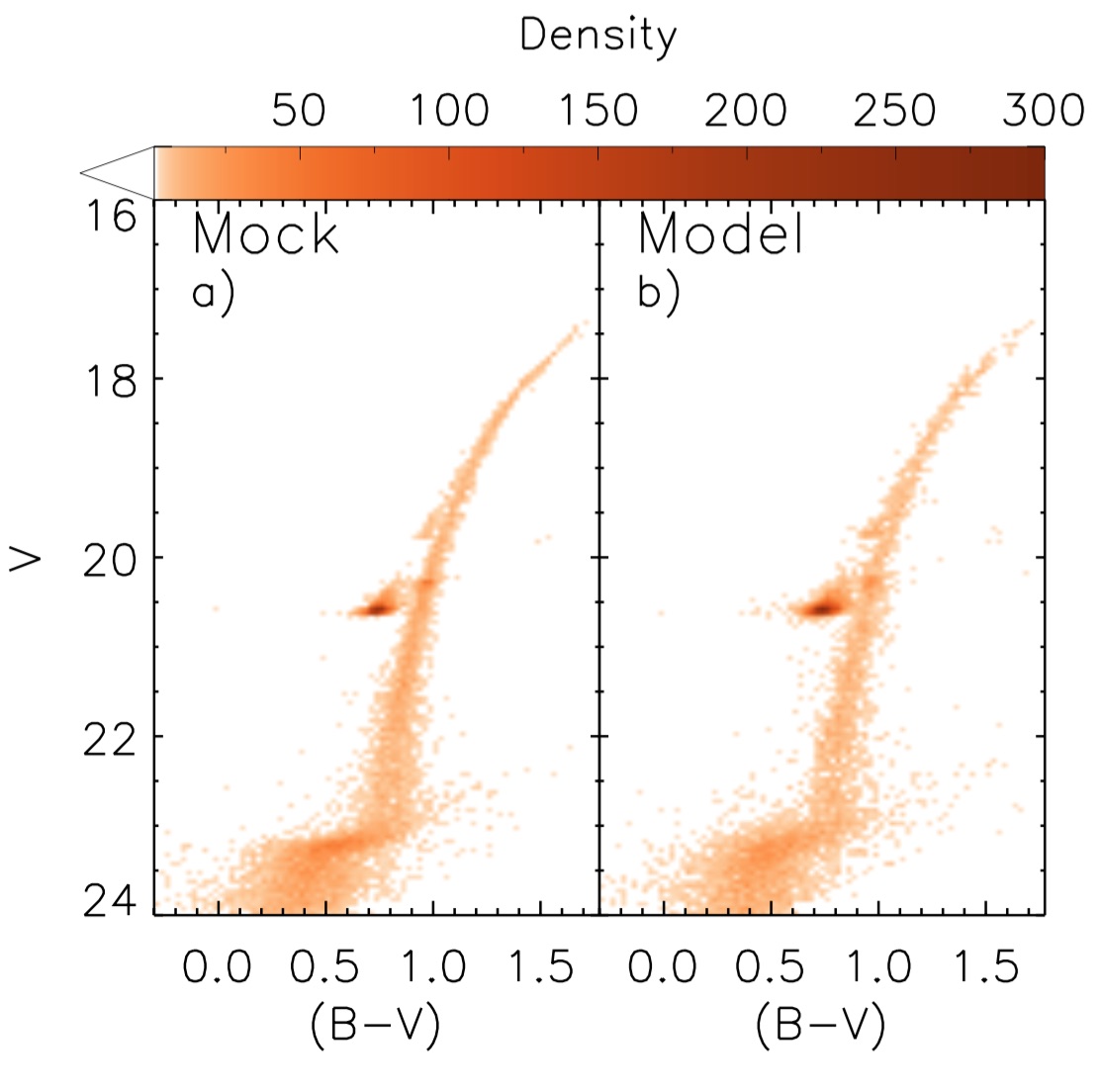}
    \caption{Same as Fig.~\ref{fig:CMDomp} but for our younger and metal richer SSP (t = 8 Gyr, [Fe/H] = -1.0 dex.)}
    \label{fig:CMDymr}
\end{figure}

\begin{figure}
	 \subfloat[][]
	{\includegraphics[width=\columnwidth]{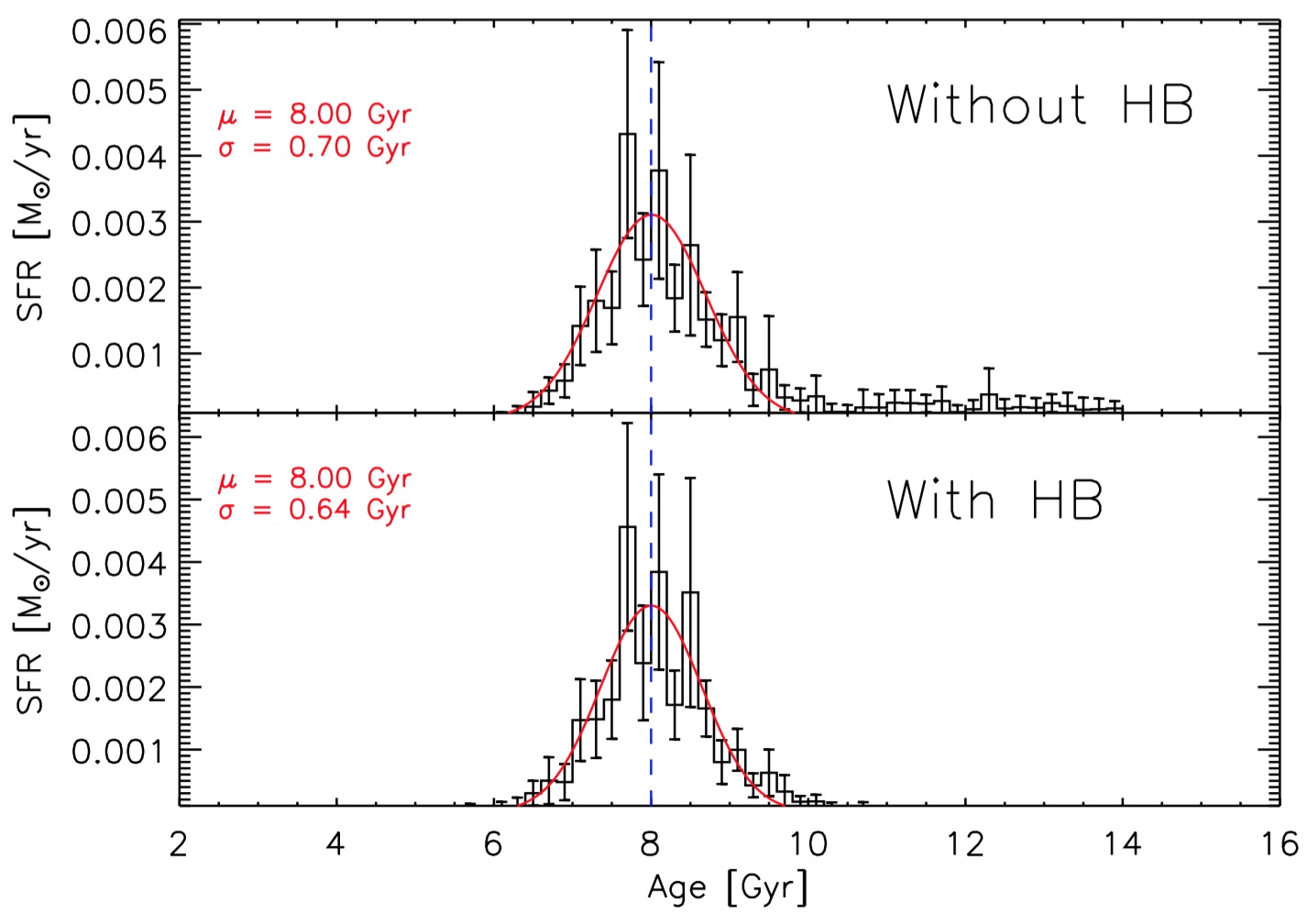}\label{fig:sfrymr}} \quad
	        \subfloat[][]
	{\includegraphics[width=\columnwidth]{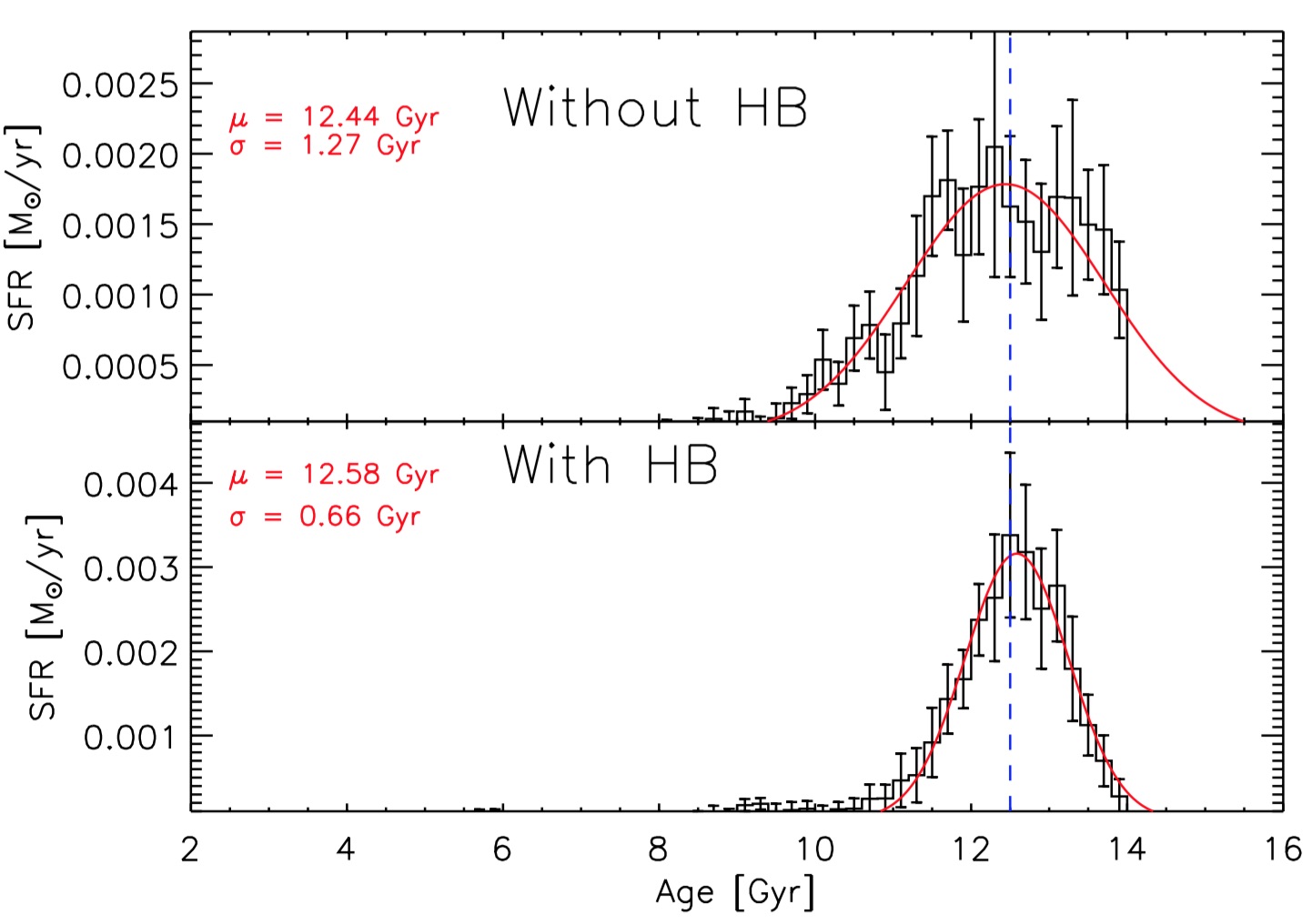} \label{fig:sfromp}}
    \caption{The comparison between the SFHs recovered with and without the HB modelling. The red solid lines are the Gaussian functions that best fit the SFHs. The blue dashed lines mark the true age of the stellar populations. \textit{a)} Solutions for the young metal rich SSP. \textit{b)} Solutions for the old metal poor SSP.}
    \label{fig:SFHSSP}
\end{figure}
In practice, we need a functional form for the integrated RGB mass loss. Recent studies of globular clusters, have suggested a linear relationship between the total amount of mass lost on the RGB and the cluster metallicity \citep{Gratton10,Origlia14}. Within the error bars, the two results agree and are also compatible with the mass loss inferred by \citet{Salaris13} for the Sculptor dSph. For this reason, we parametrise the integrated RGB mass loss as a linear function of the stellar population metallicity, and hence describe it with two parameters. In principle, the age of the stellar population can also have an effect on the RGB mass loss. However, there is little observational evidence on the importance of this parameter. The analysis of \citet{Salaris13} suggests that age has a minor effect on the mass loss of old stellar populations. For this reason we assume that the total RGB mass loss depends solely on the metallicity of the stellar population. It is possible that this parametrisation may be inappropriate when modelling younger stellar populations, as they have not been considered by these observational studies and for which the RGB evolutionary timescale (likely related to the total mass loss) is significantly affected by the stellar mass or, in other words, age. It should be noted that assumptions about the mass loss functional form only affect the link between the global CMD models and their related HB models.

\begin{figure}
	\includegraphics[width=\columnwidth]{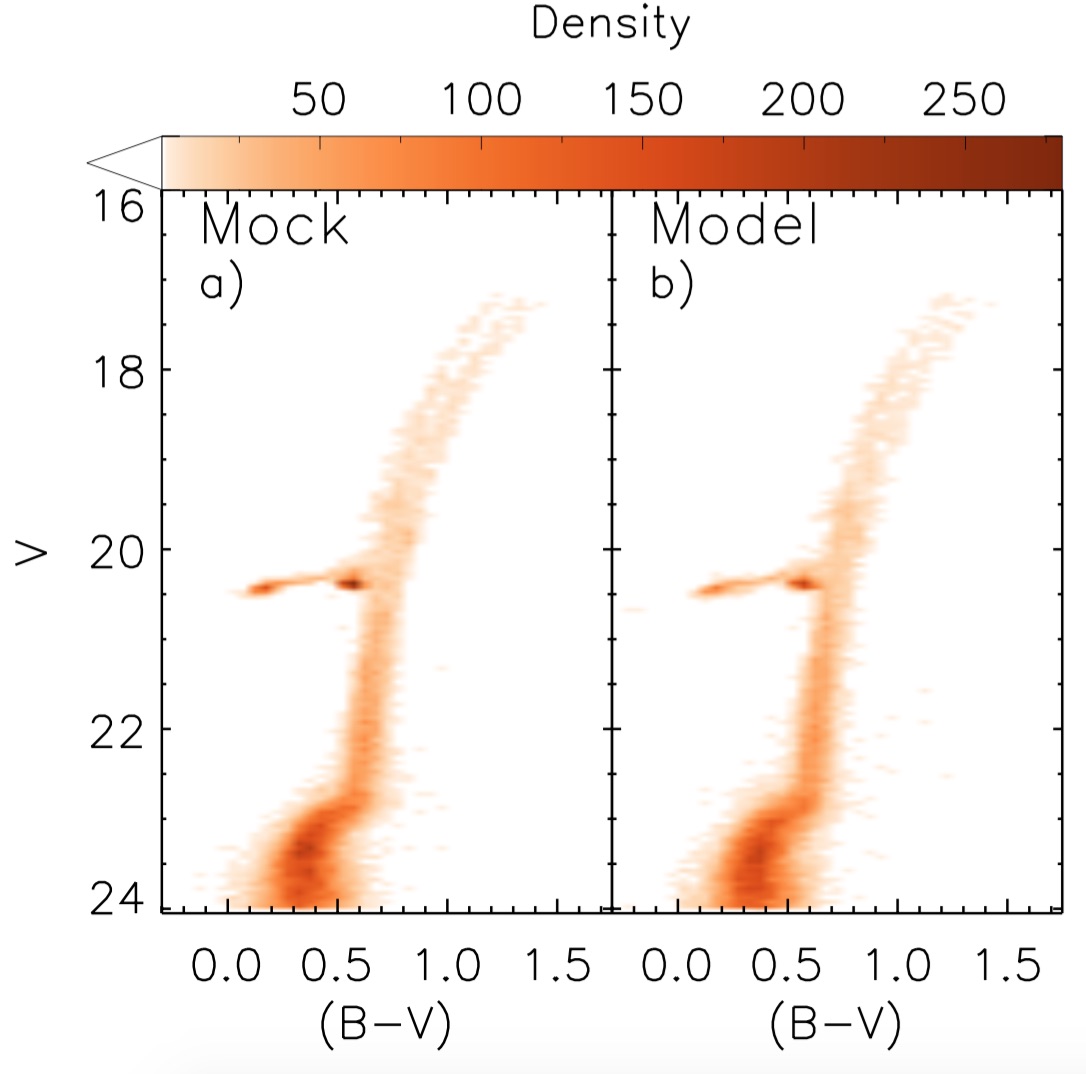}
    \caption{\textit{a)} the mock (B - V) vs V Hess diagram of our galaxy with a bursty SFH. \textit{b)} The best fit Hess diagram recovered with {\small MORGOTH}.}
    \label{fig:CMDbur}
\end{figure}

\subsection{Solving for the SFH}

With the set of models and the matching procedure described above, we can estimate the SFH of a resolved stellar population using all the evolutionary phases that are typically observed in old and intermediate age stellar populations.

\begin{figure*}
\centering
        \subfloat[][]
	{\includegraphics[width=.4\linewidth]{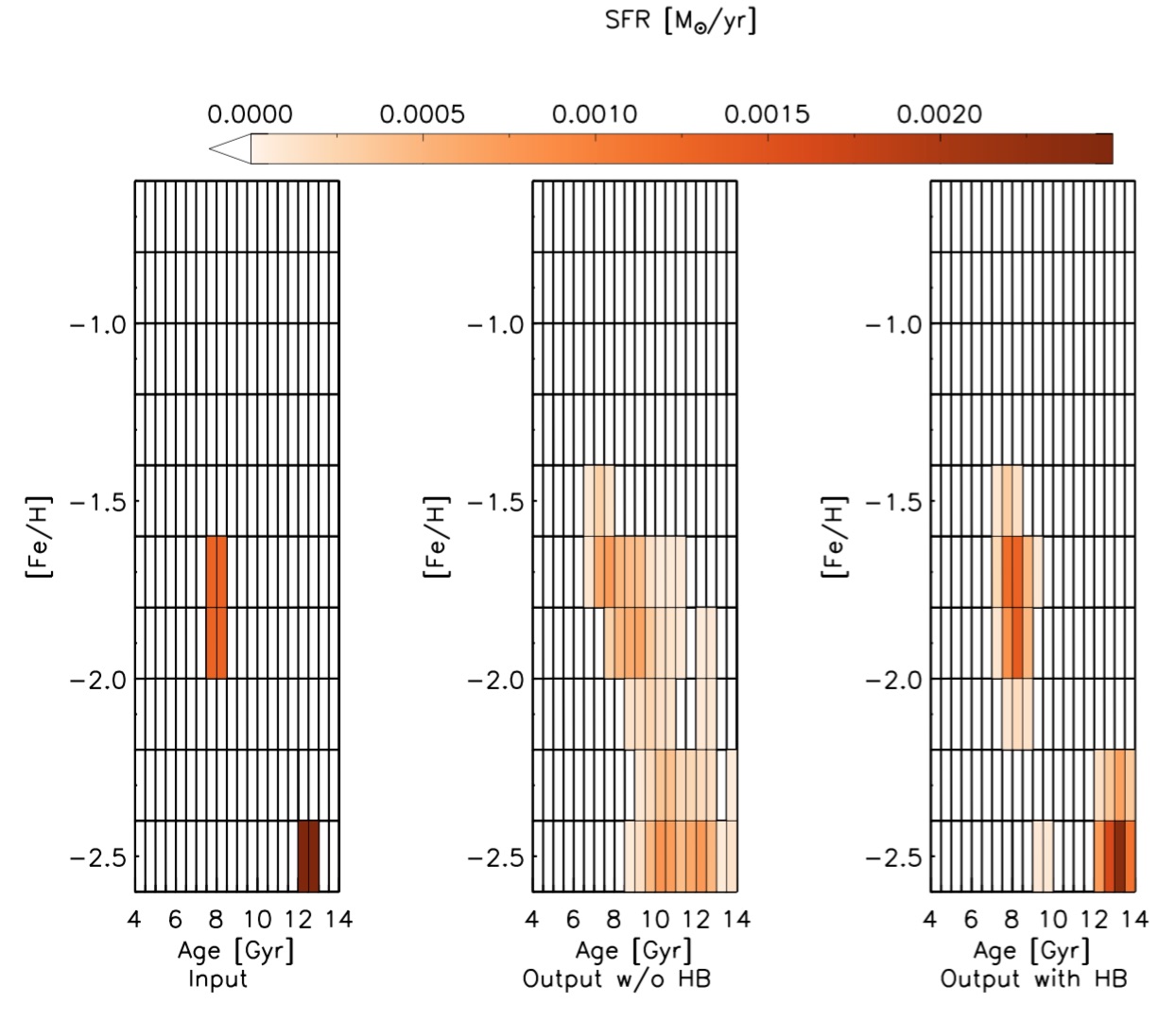} \label{fig:SFHbur}} \quad
	 \subfloat[][]{\raisebox{-48ex}
	{\includegraphics[width=.4\linewidth]{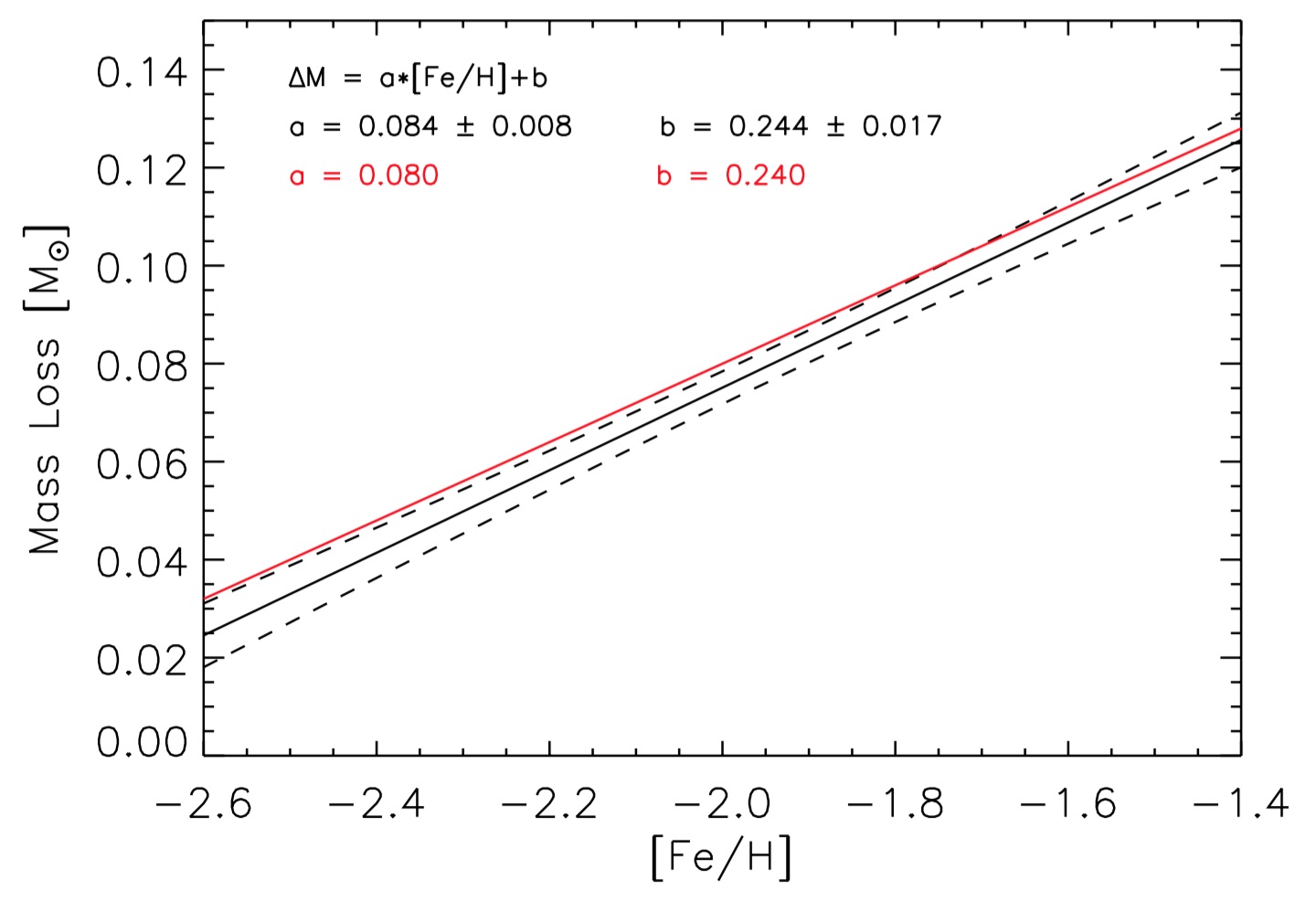} \label{fig:MLbur}}} \quad
		 \subfloat[][]
	{\includegraphics[width=.4\linewidth]{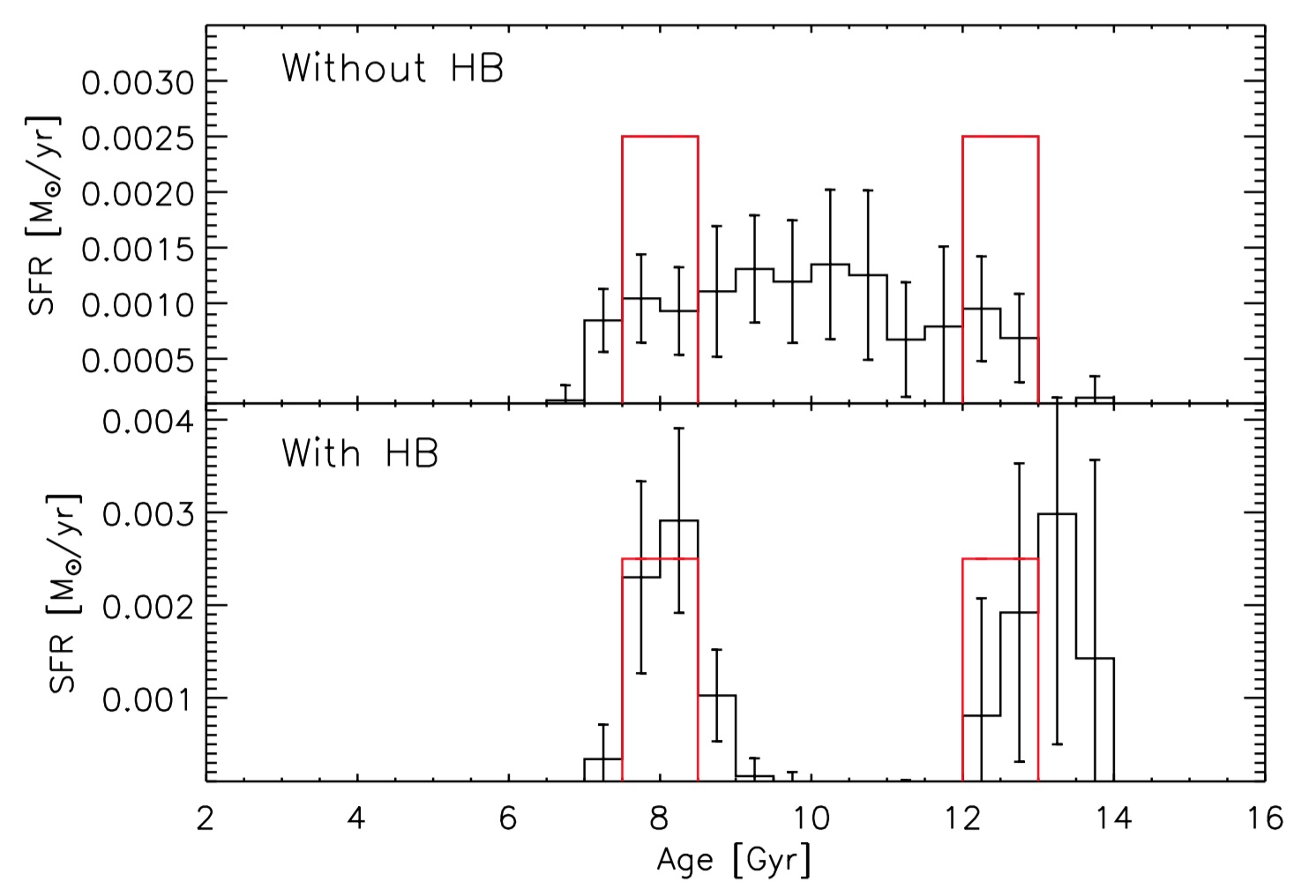}} \quad
		 \subfloat[][]{\raisebox{-39ex}
	{\includegraphics[width=.4\linewidth]{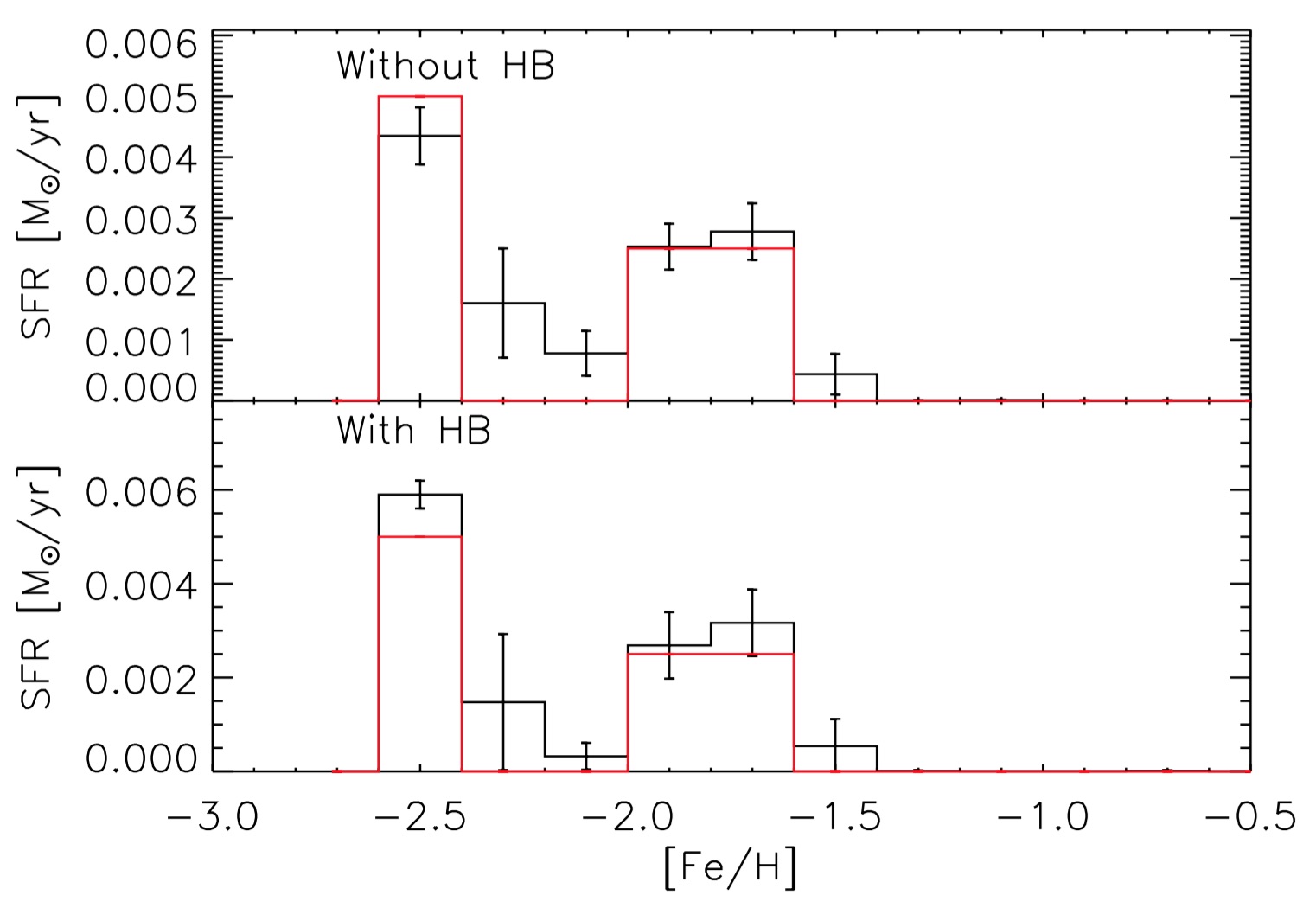}}}
    \caption{The best fit solutions for our bursty galaxy, compared with the real values. \textit{a)} The SFH in the age-metallicity plane. The three panels are for our input SFH, the one recovered with excluding the HB and the one recovered modelling also the HB (left, centre, right, respectively). \textit{b)} The recovered RGB mass loss as a function of metallicity (black solid line). The red solid line is the input mass loss prescription. The dotted black lines delimit the 1-$\sigma$ confidence region of our solution. \textit{c)} The star formation rate as a function of time, recovered with and without the HB modelling. The red solid lines represent the input values. \textit{d)} The star formation rate as a function of metallicity, recovered with and without the HB modelling. The red solid lines represent the input values.}
    \label{fig:bursty}
\end{figure*}

The main difference with previous CMD modelling approaches is the RGB mass loss and how it is treated. One way to proceed is to make an assumption about the mass loss parameters and to keep them fixed. The adopted mass loss is then used to match the HB models with the rest of the CMDs and the SFH computation proceeds as in {\small TALOS}.

Another approach is to leave the mass loss parameters free and explore them in determining the best fit SFH solution. On the HB, the age of the stellar population and the RGB mass loss are degenerate, as different combinations of these two variables can lead to the same mass on the HB. If an independent age indicator is available (for example the MSTOs), then this degeneracy can be broken and both the age and the mass loss can be determined. The HB models and the rest of the CMDs are rematched every time the mass loss is changed. This process continues until the combination of SFH and mass loss parameters that minimises $\chi^2_P$ is found.

The simplest approach would be to solve for the mass loss parameters and the SFH at the same time. Unfortunately, this creates local minima, meaning that analytical optimization techniques, such as the conjugate gradient descent used in {\small TALOS}, do not converge. Alternatively, heuristic methods, such as Markov-Chain Monte Carlo or genetic algorithms, are very slow, given the high dimensionality of the problem, introduced by the many CMD models.

As explained in \citet{Dolphin13}, if one or more parameters affects our CMD models, we assume that the likelihood of a given parameter combination is proportional to the maximum SFH likelihood (hence the minimum $\chi^2_P$) that is obtained with that combination. We, thus, explore the mass loss parameter space using an amoeba minimiser \citep{NelderMead65} until we find the mass loss prescription that produces the minimum $\chi^2_P$. We chose the amoeba minimisation because of its speed, compared to other heuristic methods. The topology of our $\chi^2_P$ surface is such that this method efficiently converges without getting trapped in the local minima, because the scale and the depth of these minima, which arise because of our discrete HB grid, are much smaller than the large scale structure of the $\chi^2_P$ surface. This means that the minimiser is insensitive to the presence of these features until it is very close to the global minimum (on a scale of the order of the HB mass sampling). This hybrid analytical-heuristic approach allows us to overcome the local minima introduced by the mass loss while exploiting the speed of the SFH optimisation.

\subsection{Uncertainty estimation}
\label{errors}
Once the preferred SFH solution has been found, we estimate the uncertainties on our solution. New solutions are calculated for different CMD and parameter space samplings, as described in \S\ref{talos}. For each of these, we generate additional CMDs using the \textit{Poisson statistics} criterion, described in \citet{Aparicio09}. For each bin $i$ of the observed CMD, we replace the measured stellar count $n_i$ with a new value, drawn from a Poisson distribution with  expectation value $\lambda=n_i$. These new CMDs are used to estimate the uncertainties arising from data sampling.

The way mass loss is incorporated into the uncertainty estimation depends on whether the mass loss parameters were fixed. If the mass loss is left as a free parameter, a separate measurement is made for each SFH solution. Similarly to the SFH analysis, the final mass loss is the average of all the individual solutions and the uncertainties on the parameters are taken from the standard deviation of the solutions. To prevent solutions that produce a bad CMD fit to contaminate our final measurement, we apply a 5.19 median absolute deviation cut on the mass loss parameter distribution, eliminating strong outliers. This corresponds roughly to a 3.5 sigma clipping.

If the mass loss parameters are fixed, the uncertainties on these parameters are taken into account when computing uncertainties on the SFH. The entire process of SFH evaluation and the uncertainty estimation is repeated several times. Each time the mass loss parameters are extracted from a gaussian distribution, according to their uncertainties. The final SFH will be the one corresponding to the mass loss that gives us the minimum $\chi^2_P$, whereas the scatter of the entire set of solutions will provide the confidence interval for our SFH.
 
The uncertainties on the individual star formation rates are highly correlated. For this reason, {\small MORGOTH} also estimates the covariance between each pair of star formation rates. This information is needed to treat the SFH uncertainties properly, as varying the star formation rates independently could lead to unphysical solutions that, for example, do not match the mass of the galaxy.

\section{Testing the method}
\label{test}

It is necessary, with a new analysis procedure, to benchmark it on science cases for which the expected solution is known. The standard approach is to use synthetic CMDs with known SFH to emulate typical observed examples. Of course, these tests on mock observations represent an idealized situation and, while they are very useful to assess the validity of the adopted methodology, they are free from many of the systematics that affect real data. These include uncertainties in the modelling parameters (such as distance, reddening and alpha enhancement), mismatch with theoretical models and poor characterisation of the observational effects. Thus we also apply the method to a well studied object with a previous analysis, including the HB modelling. This has the advantage of working with ``real'' data, testing that the method is robust against the typical uncertainties of observations. However, in this case, we can only make a relative comparison with previous studies.

In this section and in the next, we are going to present tests on synthetic datasets of different complexity, and on the well studied Local Group galaxy Sculptor.
\subsection{Synthetic tests on simple stellar populations}
\label{SSP}

\begin{figure}
	\includegraphics[width=\columnwidth]{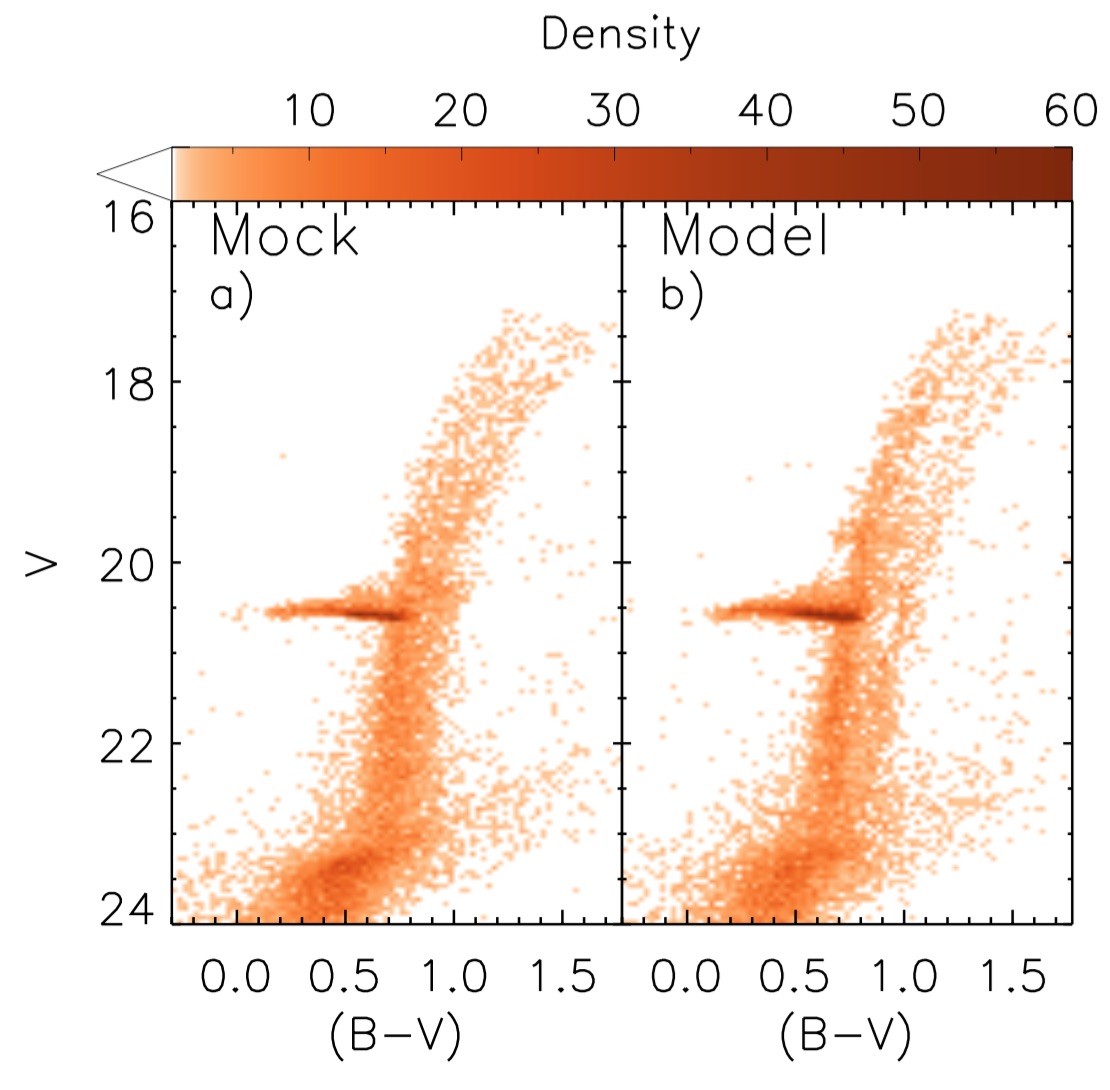}
    \caption{\textit{a)} The mock (B - V) vs V Hess diagram of our galaxy with a continuous SFH. \textit{b)} The best fit Hess diagram recovered with {\small MORGOTH}.}
    \label{fig:CMD2dex}
\end{figure}

\begin{figure*}
        \subfloat[][]
	{\includegraphics[width=.4\linewidth]{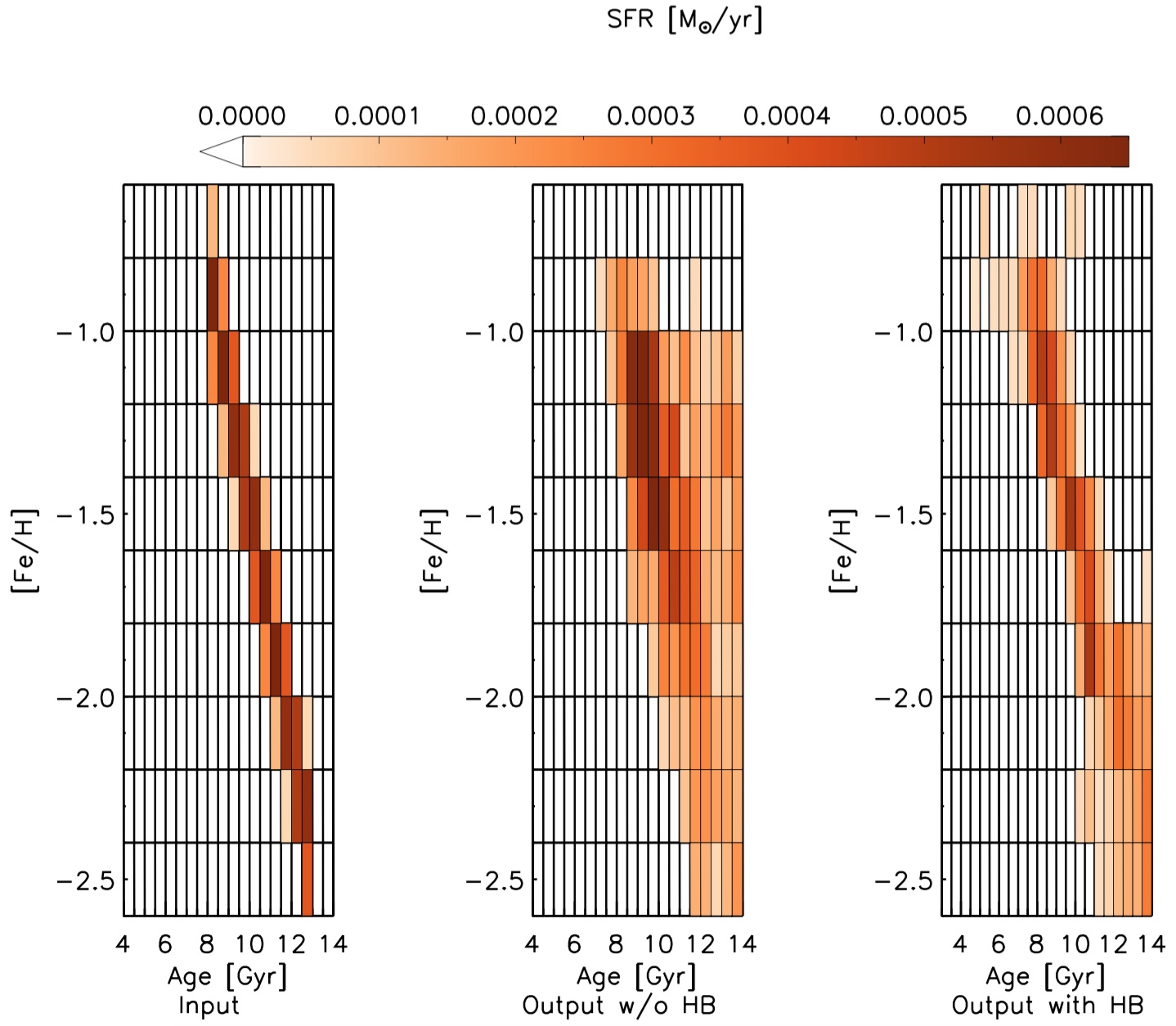} \label{fig:SFH2}} \quad
	 \subfloat[][]{\raisebox{-50ex}
	{\includegraphics[width=.4\linewidth]{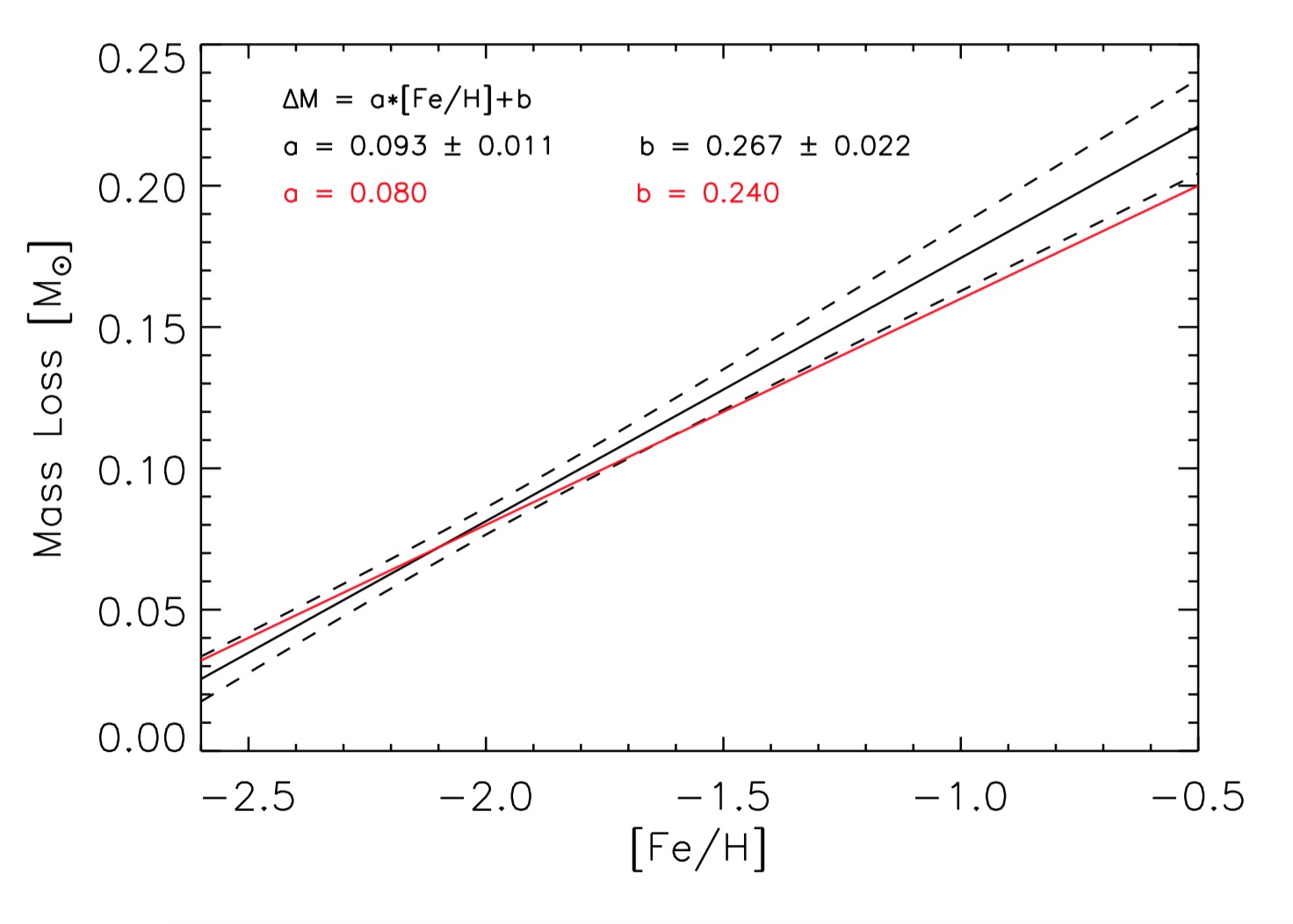}}} \quad
		 \subfloat[][]
	{\includegraphics[width=.4\linewidth]{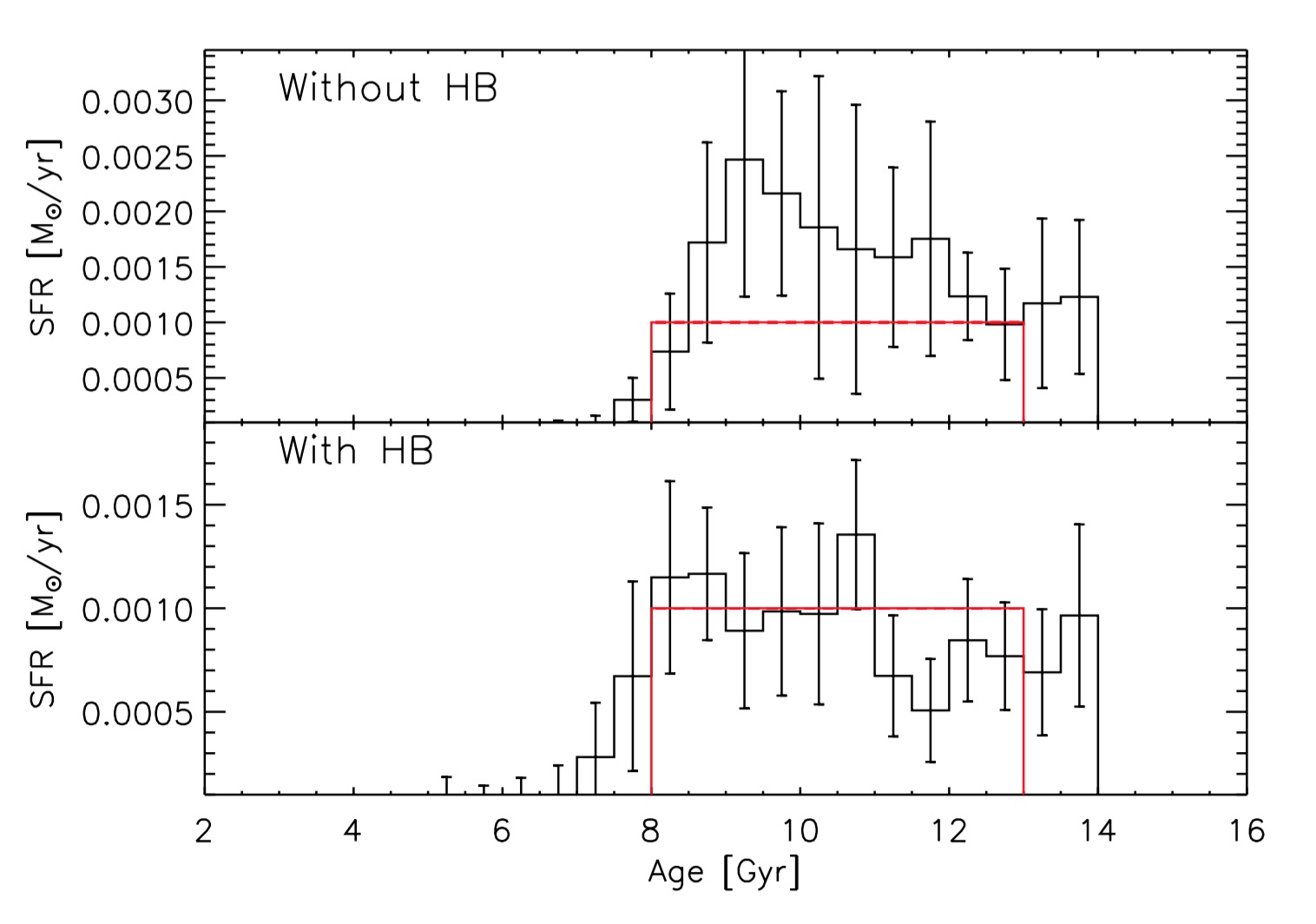}} \quad
		 \subfloat[][]{\raisebox{-39ex}
	{\includegraphics[width=.4\linewidth]{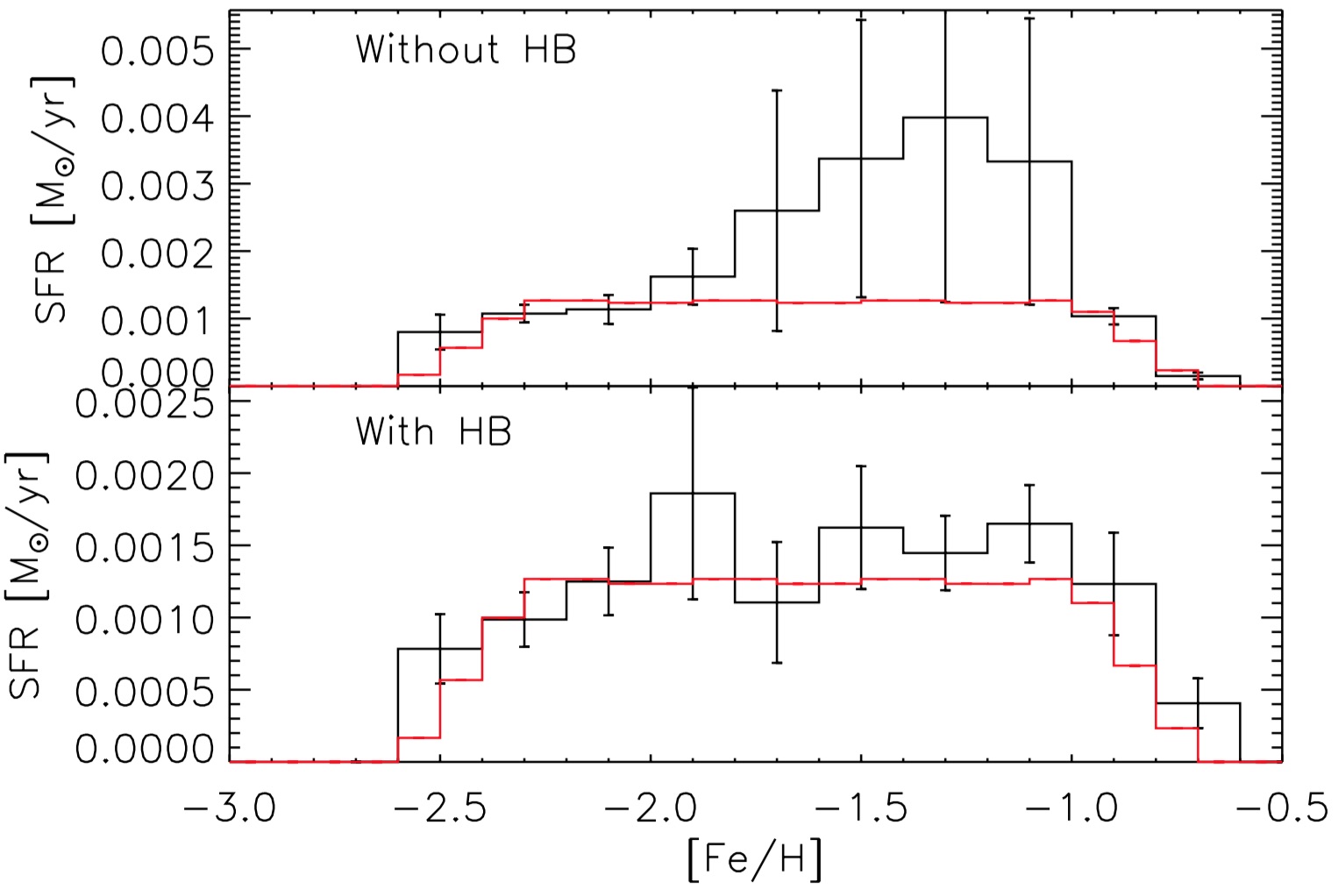}}}
    \caption{The best fit solutions for our galaxy with continuous SFH, compared with the real values. \textit{a)} The SFH in the age-metallicity plane. The three panels are for our input SFH, the one recovered with excluding the HB and the one recovered modelling also the HB (left, centre, right, respectively). \textit{b)} The recovered RGB mass loss as a function of metallicity (black solid line). The red solid line is the input mass loss prescription. The dotted black lines delimit the 1-$\sigma$ confidence region of our solution. \textit{c)} The star formation rate as a function of time, recovered with and without the HB modelling. The red solid lines represent the input values. \textit{d)} The star formation rate as a function of metallicity, recovered with and without the HB modelling. The red solid lines represent the input values.}
    \label{fig:2dex}
\end{figure*}

The performance of a SFH modelling technique should pass the basic test applied to simple stellar populations \citep{Hidalgo11,deBoer12}. A simple stellar population is defined to have a single age and metallicity (or an age and metallicity spread negligible compared to the resolution of either). SFH determinations are typically broader than the input, due to a combination of observational effects and methodological limitations \citep{Hidalgo11,Aparicio16}. Measuring this effect for a simple stellar population gives an estimate of the intrinsic time resolution of the method.

Globular clusters are typically characterized by very small spreads in age and metallicity, and could be interesting targets to test {\small MORGOTH} on real simple stellar populations. However, globular clusters are also known to have substantial spreads in certain light element abundances, including helium \citep[see, e.g.,][and references therein]{Gratton12}. These spreads have a strong influence on the morphology of the HB, which thus cannot be straightforwardly linked to the SFH of the system. The field stellar populations of nearby resolved galaxies show no sign of these abundance patterns \citep{Shetrone03, Geisler07, Fabrizio15}, but they are characterized by an extended SFH. For this reasons, we can only perform simple tests on synthetic data.

We created a synthetic CMD for two simple stellar populations. One with an age of 12.5 Gyr and a metalicity, [Fe/H] = -2.5 dex, and the other with an age of  8 Gyr old and [Fe/H] = -1.0 dex. We chose these values to span the range of age and metallicity typically observed in Local Group dSphs, such as Sculptor. This is also the region of the age-metallicity parameter space where stellar populations exhibit extended HBs. We emulated realistic observational effects for a dust free stellar system at a distance of 100 kpc, using the photometric errors and the artificial star tests of the Sculptor dSph CTIO observations, presented in \S~\ref{Sculptor}. We made these tests for a fixed mass loss law, taken from \citet{Origlia14}. The resultant CMDs are shown, as Hess maps, in Fig.~\ref{fig:CMDomp} and Fig.~\ref{fig:CMDymr}, and the recovered SFHs are shown in Fig.~\ref{fig:SFHSSP}.

We first determined the SFH without the HB, as done by {\small TALOS} and similar methods. The recovered SFHs are well fit by a Gaussian function with standard deviation of 700 Myr, for the young population, and 1.3 Gyr, for the old population. These values are in line with what found by other studies \citep[e.g.,][]{Hidalgo11,deBoer12}.


Fig.~\ref{fig:SFHSSP} shows that the inclusion of the HB leads to an improvement in the precision of the recovered SFHs, which now have a standard deviation of 650 Myr, compared to the modelling without the HB. This improvement is mild for the younger population (Fig.~\ref{fig:sfrymr}), where the bright MSTO stars allow a fairly precise SFH. However, for the older population (Fig.~\ref{fig:sfromp}), the MSTO is fainter and characterized by a slower morphology evolution with age. In this case, the inclusion of the HB, whose morphology is very sensitive to stellar mass, significantly improves the age measurement, increasing the precision by a factor of $\sim 2$.

Obviously, these experiments with synthetic data represent an ideal situation and give us the maximum theoretical precision that this approach can reach. In reality, any discrepancy between the synthetic models and a real galaxy, such as distance, reddening or stellar evolutionary track mismatches or uncertainties, will degrade our model-data comparison.

 Among these systematics, a major role is played by mass loss, which was assumed to be known in our modelling. Any mismatch between the predicted and true mass loss will bias the age determination of the stellar population, because of the degeneracy of these two parameters on the HB. The importance of this effect is given by the derivative of the RGB tip mass over time, at a given age. For a 12.5 Gyr old population this amounts to $\sim 0.02\, M_{\odot} Gyr^{-1}$, with a slight dependency on metallicity, while for a 8 Gyr old population it is $\sim 0.04\, M_{\odot} Gyr^{-1}$. This translates to an effect on the SFH of 500 Myr and 250 Myr, respectively, for each $0.01\, M_{\odot}$ of difference in the mass loss.

\subsection{Moving beyond single age synthetic models}

The simple stellar population tests we performed confirm the potential of HB morphology to add additional independent constraints to standard SFH analysis and significantly improve age resolution at old times. A clear limitation of our simple stellar population experiments is the assumption that the RGB mass loss is known. Another shortcoming is the very simple SFH adopted. Real galaxies will typically have much more complex SFHs, spanning a range of age and metallicity, reflected in the morphology of their CMDs.

To address both these issues we carried out additional experiments with synthetic data to simulate more realistic conditions. We created synthetic CMDs for two galaxies with different SFHs. We used the same observational conditions as in \S~\ref{SSP} (distance, photometric uncertainties, completeness, etc.). We then run {\small MORGOTH} to recover the SFH, as well as the RGB mass loss, for which no assumption was made, except a linear dependence on [Fe/H].

In the first experiment we simulate a galaxy with two short bursts of star formation, separated by $\sim 4$ Gyr in age and $\sim 0.7$ dex in metallicity. The mock CMD of this galaxy is shown in Fig.~\hyperref[fig:CMDbur]{5a}. Fig.~\hyperref[fig:CMDbur]{5b} shows our best fit CMD. It can be seen that our best fit is a realistic representation of the observed CMD, also for the HB morphology. 

The four panels of Fig.~\ref{fig:bursty}, show the original inputs of our mock galaxy, along with the recovered results. As a reference, we also fit the SFH in the classical way, i.e., excluding the HB. As for the CMD, the recovered SFH is similar to the original input. In particular, the comparison with the reference solution demonstrates that the inclusion of the HB modelling helps to improve separating and characterising these two bursts of star formation, both in age and metallicity. Fig.~\ref{fig:MLbur} shows the mass loss relation that we recover, compared to that used as input. The simultaneous modelling of the HB and the MSTO region efficiently recovers the RGB mass loss prescription, within one sigma of the original.

We perform the same experiment on a second mock CMD (Fig.~\hyperref[fig:CMD2dex]{7a}), built using a constant star formation rate and a steady metallicity evolution (Fig.~\ref{fig:SFH2}). Again, the solutions we recover with and without the HB modelling, shown in Fig~\ref{fig:2dex}, prove that including the HB provides more realistic star formation rates, both as a function of age and metallicity. The HB morphology is well reconstructed, as demonstrated by the RGB mass loss inferred to be so closely matching the input.

These additional tests show that {\small MORGOTH} is able to efficiently reconstruct the HB morphology when modelling stellar systems with extended SFHs, even when the RGB mass loss is an unknown and is included in the fitting. Furthermore, the improved accuracy given by the HB constraints is also present in the fit of more complex SFHs. This is clearly visible in the three panels of Fig.~\ref{fig:SFHbur} and Fig.~\ref{fig:SFH2}, where the SFHs recovered with the aid of the HB are obviously improved, compared to those recovered without the HB.

\begin{figure}
	\includegraphics[width=\columnwidth]{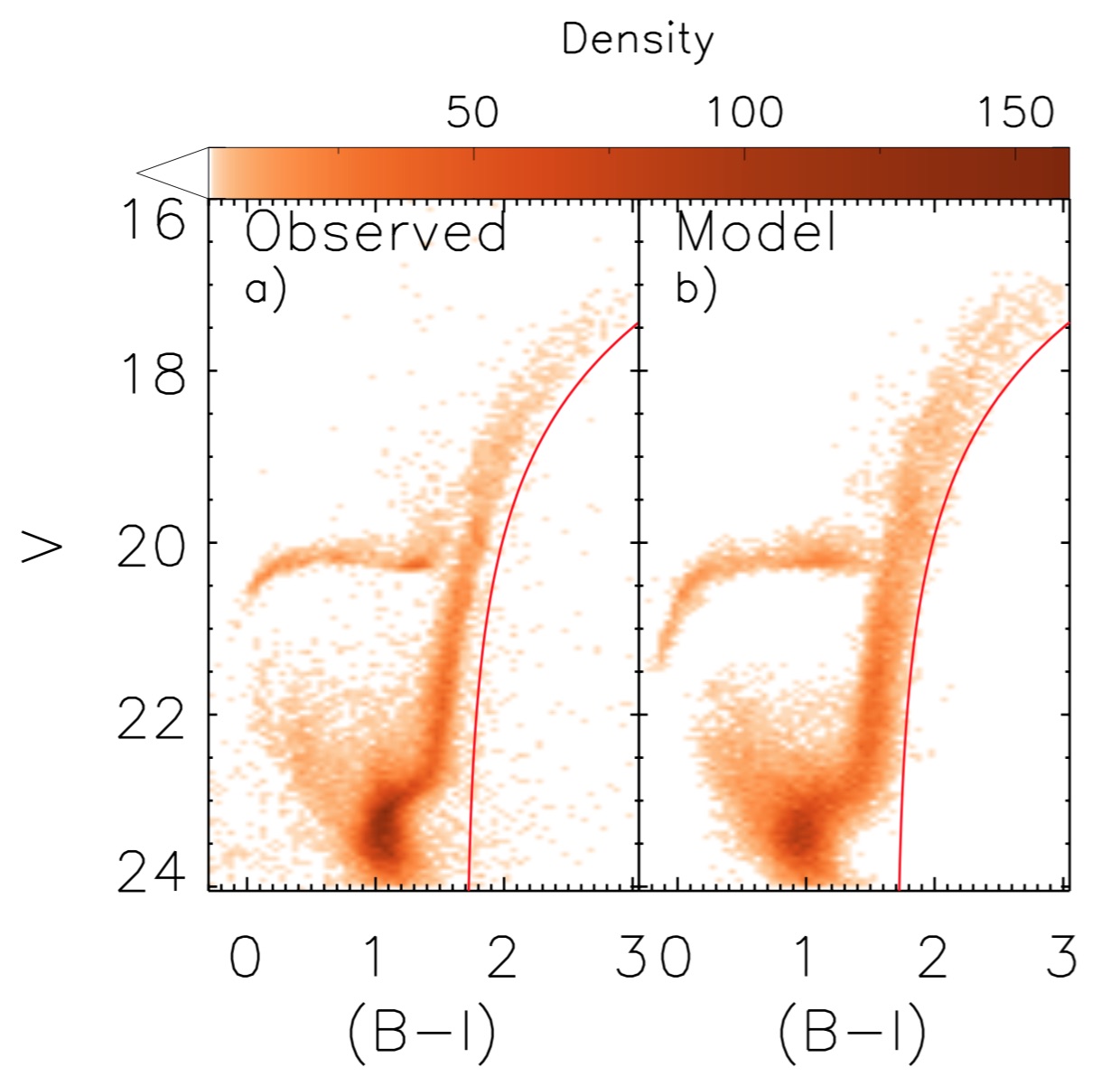}
    \caption{\textit{a)} V vs (B-I) Hess map of the Sculptor dSph. The photometric catalogue is taken from \citet{deBoer11} and it has been corrected for RR Lyrae variability. \textit{b)} Best fit Hess diagram obtained with {\small MORGOTH.} The red solid lines mark the punishment region for RGB models.}
    \label{fig:CMDscl}
\end{figure}

\section{The Sculptor dwarf spheroidal}
\label{Sculptor}

Having successfully shown the effectiveness of our method on synthetic stellar populations, we move to real data and model the resolved population of the well studied Sculptor dSph. This is a close stellar system with a relatively simple SFH, as shown by the detailed SFH from \citet{deBoer12}, using {\small TALOS}. The SFH from \citet{deBoer12} has also been used as a starting point by \citet{Salaris13}, to estimate the RGB mass loss of Sculptor. This means that both the end products of our analysis, the SFH and the RGB mass loss, have already been measured for this galaxy. As {\small TALOS} is the code from which we developed {\small MORGOTH}, this galaxy will provide a very instructive comparison to evaluate the effect of adding constraints from the HB. 

To model the CMD of Sculptor, we use the MOSAIC, CTIO 4m photometry from \citet{deBoer11}. We limit the analysis to an elliptical area in the centre of the galaxy, with an equivalent radius of 11 arcmin. This is the same region for which estimates were made of the RGB mass loss by \citet{Salaris13}.

For our analysis, we also have to account for the presence of RR Lyrae variables. When observed only in one or few epochs, these variable stars will have a wide distribution in magnitude and colour, because they are in different phases of their pulsation cycle, thus changing the HB morphology. To model the HB of Sculptor, we need to know the mean magnitude of its RR Lyrae, so that we can reconstruct the original HB morphology in the instability strip. To do that, we cross-correlated our photometric catalogue with the sample of variables from \citet{Marvaz16}. In this way, we identify most of the RR Lyrae and substitute their observed magnitudes with their intensity averaged magnitudes. The final photometric sample we obtain is shown, as a Hess map, in Fig.~\ref{fig:CMDscl}.

\begin{figure*}
\centering
        \subfloat[][]
	{\includegraphics[width=.4\linewidth]{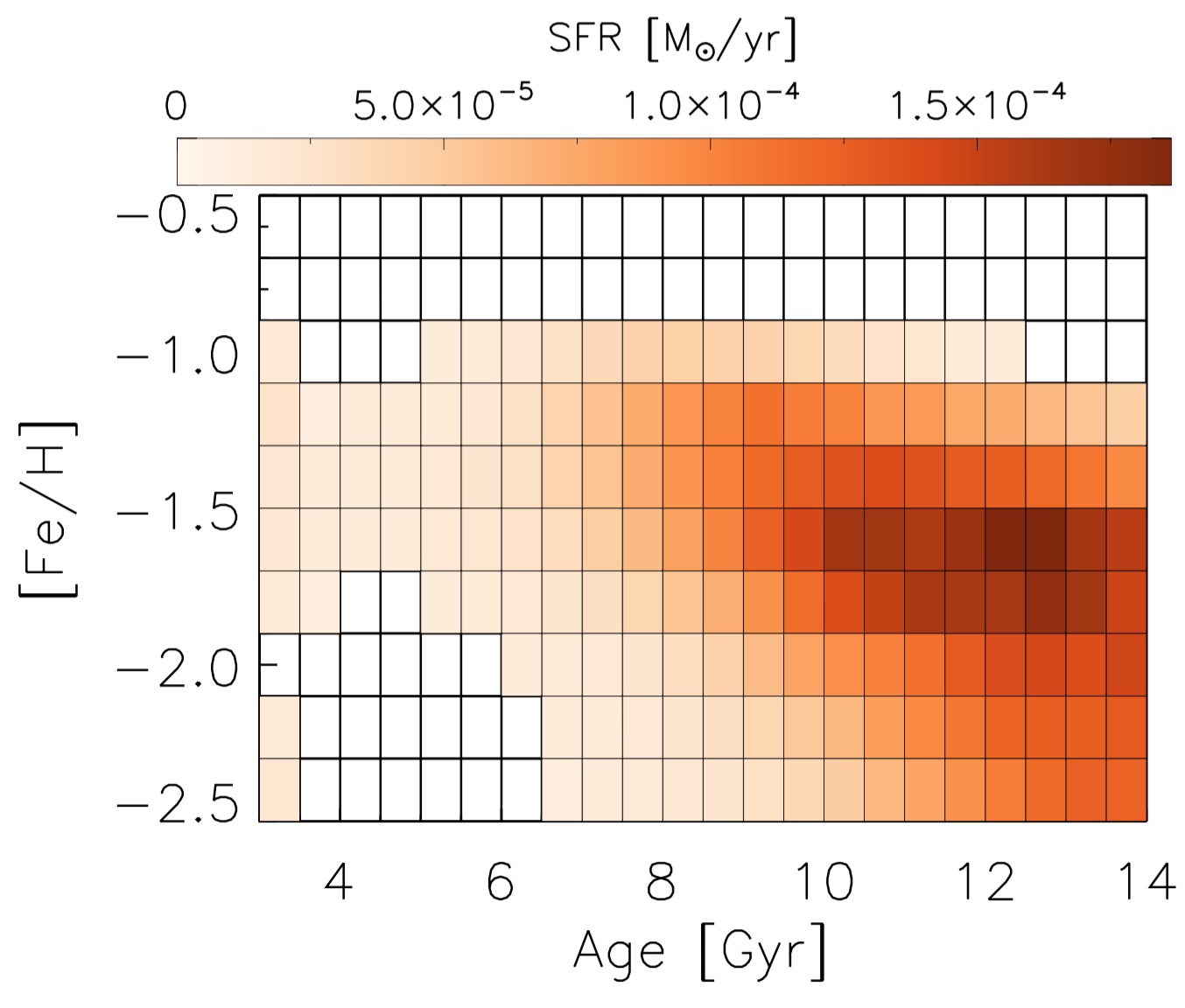} \label{fig:SFHscl}} \quad
	 \subfloat[][] {\raisebox{-48ex}
	{\includegraphics[width=.4\linewidth]{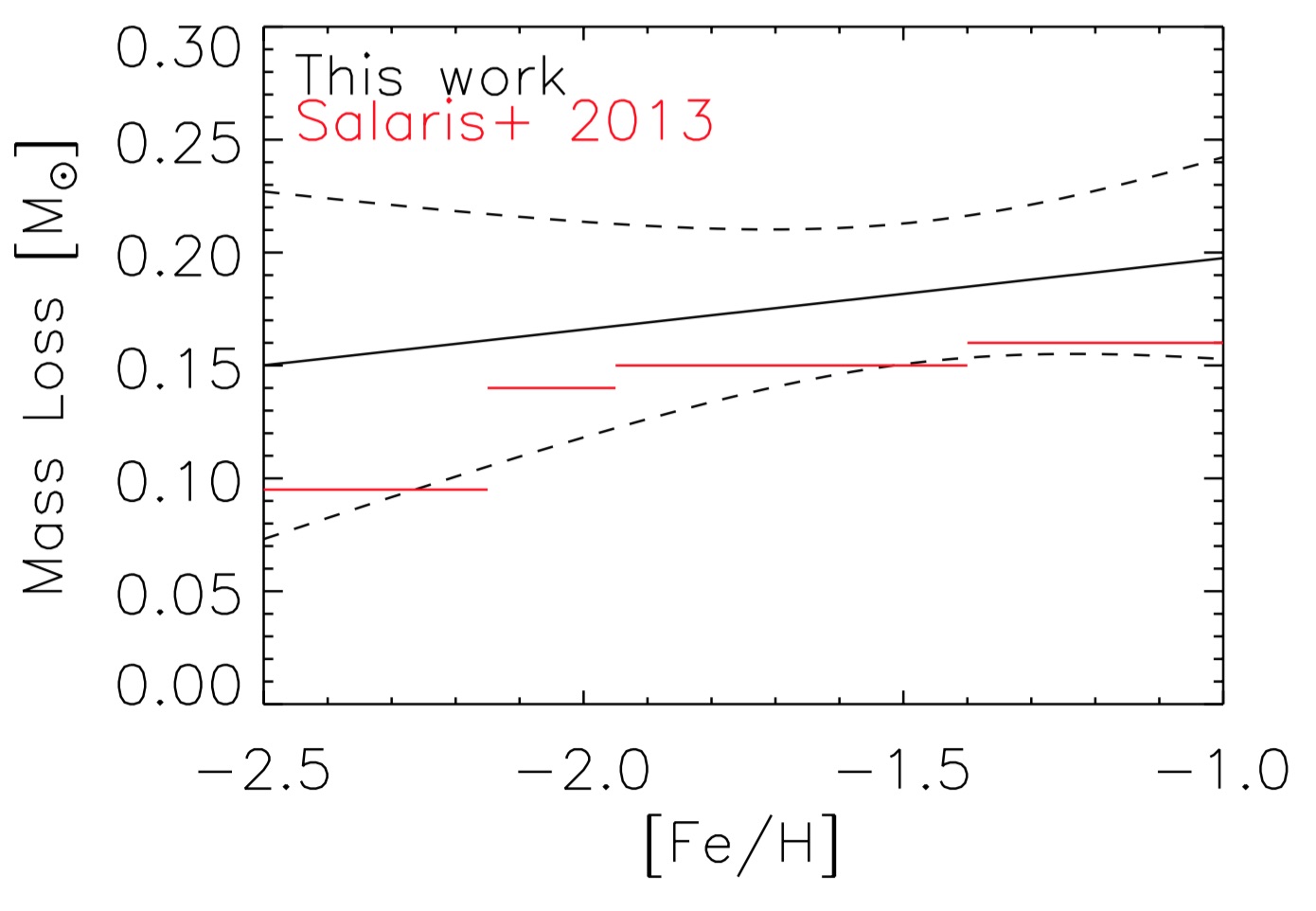}}} \quad
		 \subfloat[][]{\raisebox{-39ex}
	{\includegraphics[width=.4\linewidth]{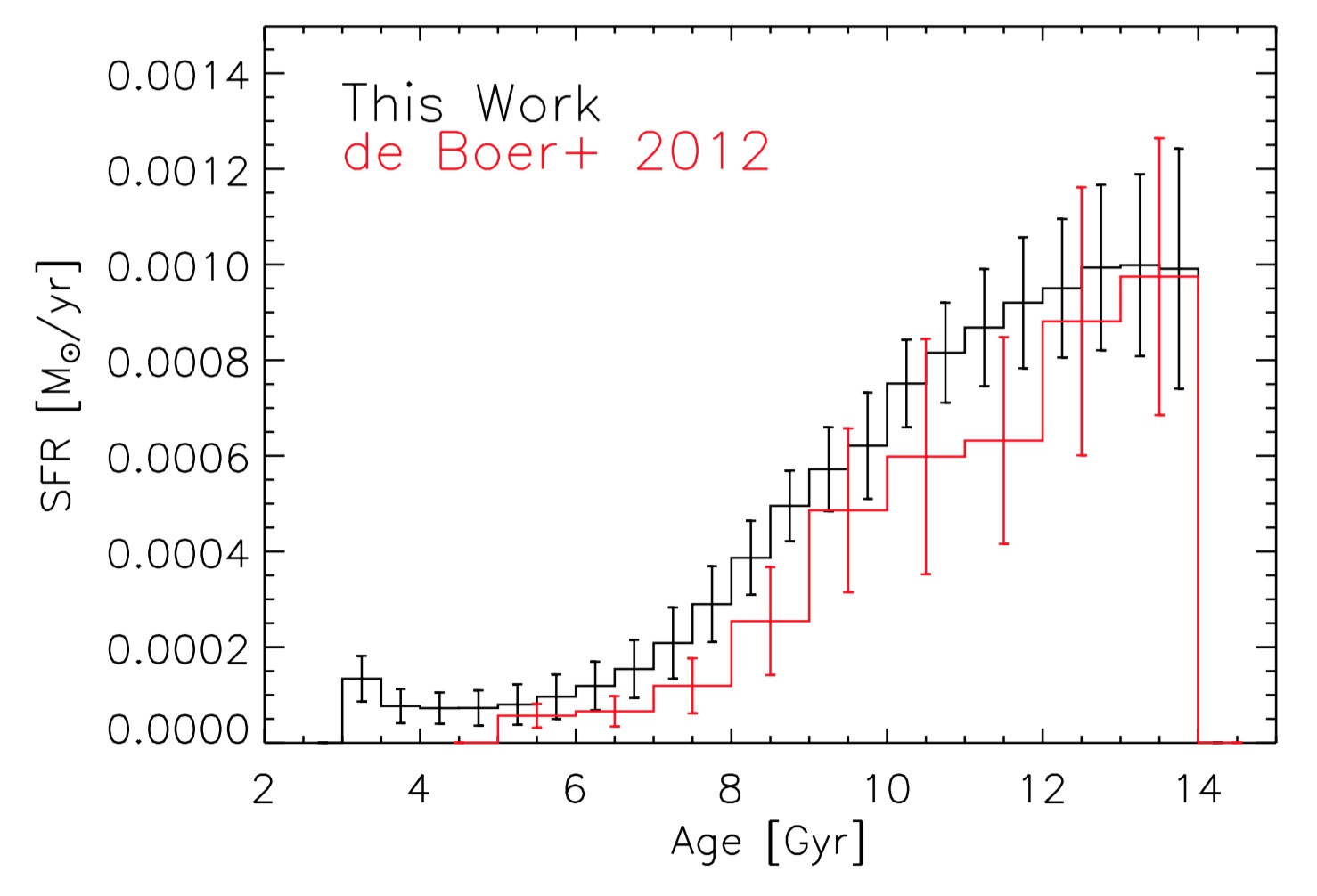}}} \quad
		 \subfloat[][]
	{\includegraphics[width=.4\linewidth]{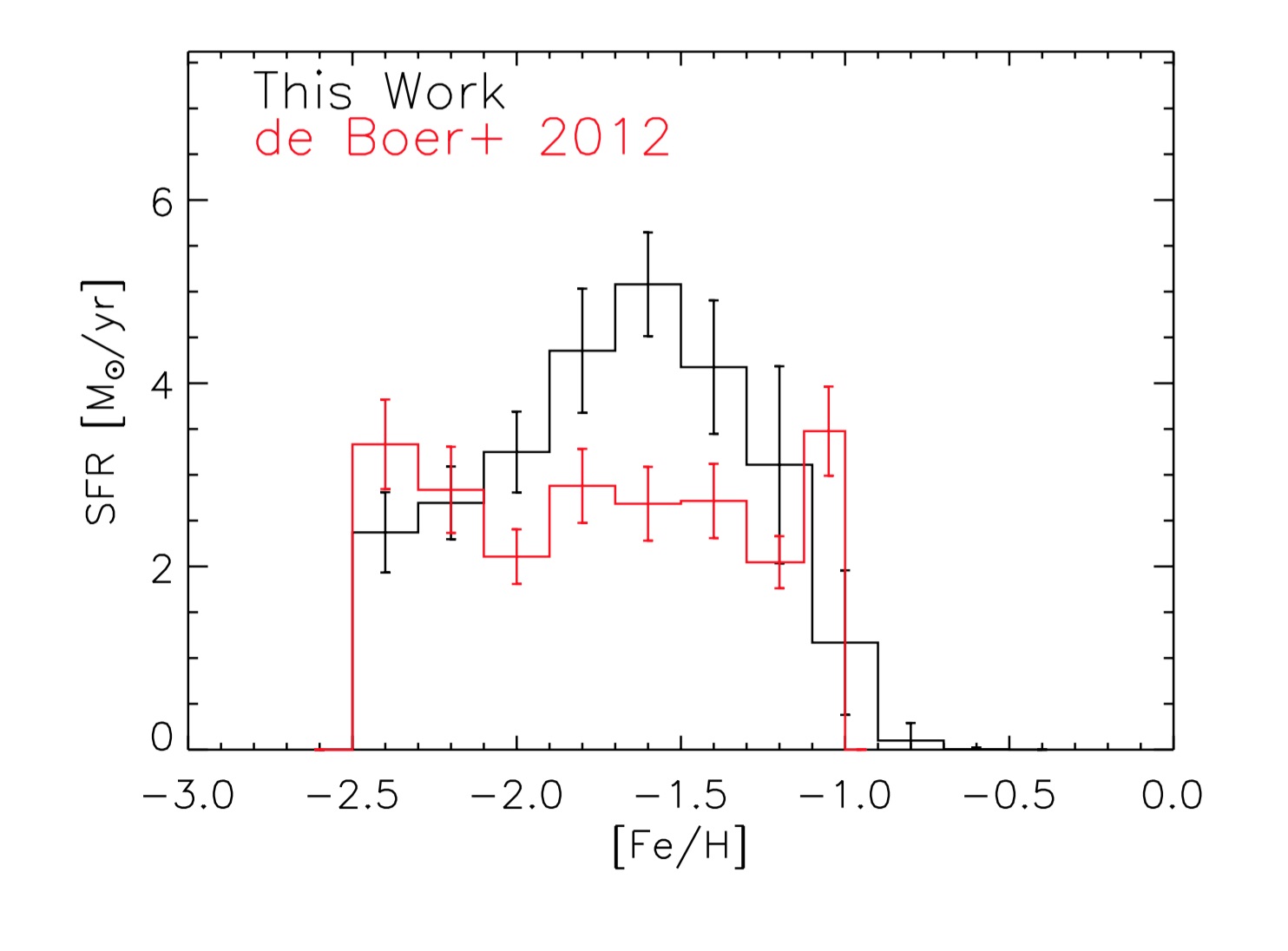}}
    \caption{\textit{a)} The age-metallicity distribution of star formation for Sculptor, recovered with {\small MORGOTH}. \textit{b)} The inferred RGB mass loss (black solid line). The dotted black lines mark the 1-$\sigma$ uncertanties of our estimate. The red solid line shows the results from \citet{Salaris13}. \textit{c)} Star formation rate as a function of cosmic time. The red histogram shows the solution of \citet{deBoer12}. \textit{d)} Star formation rate as a function of metallicity.}
    \label{fig:Scl}
\end{figure*}

We model the V vs (B-I) CMD of Sculptor, with a fine bin size (0.05 mags) to take advantage of the detailed structure of the HB. We chose this filter combination to maximize the colour extension on the HB. The model populations we use sample the age-metallicity parameter space with a step of 0.5 Gyr and 0.2 dex. Our synthetic populations follow a Kroupa IMF \citep{Kroupa01} and [$\alpha$/Fe] measured by high resolution spectroscopy \citep{Battaglia08,Tolstoy09}. We model the observational effects using both photometric errors and artificial star tests from \citet{deBoer11,deBoer12}. Finally, we adopt a distance modulus of $(m-M)_0 = 19.67$ and a value of $E(B-V) = 0.018$, as in \citet{deBoer12}.

As we are now modelling the whole CMD, in contrast to \citet{deBoer12}, we also include the full extent of the RGB. This evolutionary phase is very prominent in the CMD of an old stellar population. The detailed morphology of the RGB, especially its colour, is quite sensitive to a number of uncertain stellar model ingredients, such as the treatment of convection, the boundary conditions of the stellar model and the bolometric corrections of the cool stellar atmospheres. We compensate for these uncertainties by creating a punishment region in our CMD, as shown in Fig.~\ref{fig:CMDscl}. Every stellar model that falls on the red side of the line is heavily penalised in our fit.

\begin{figure}
	\includegraphics[width=\columnwidth]{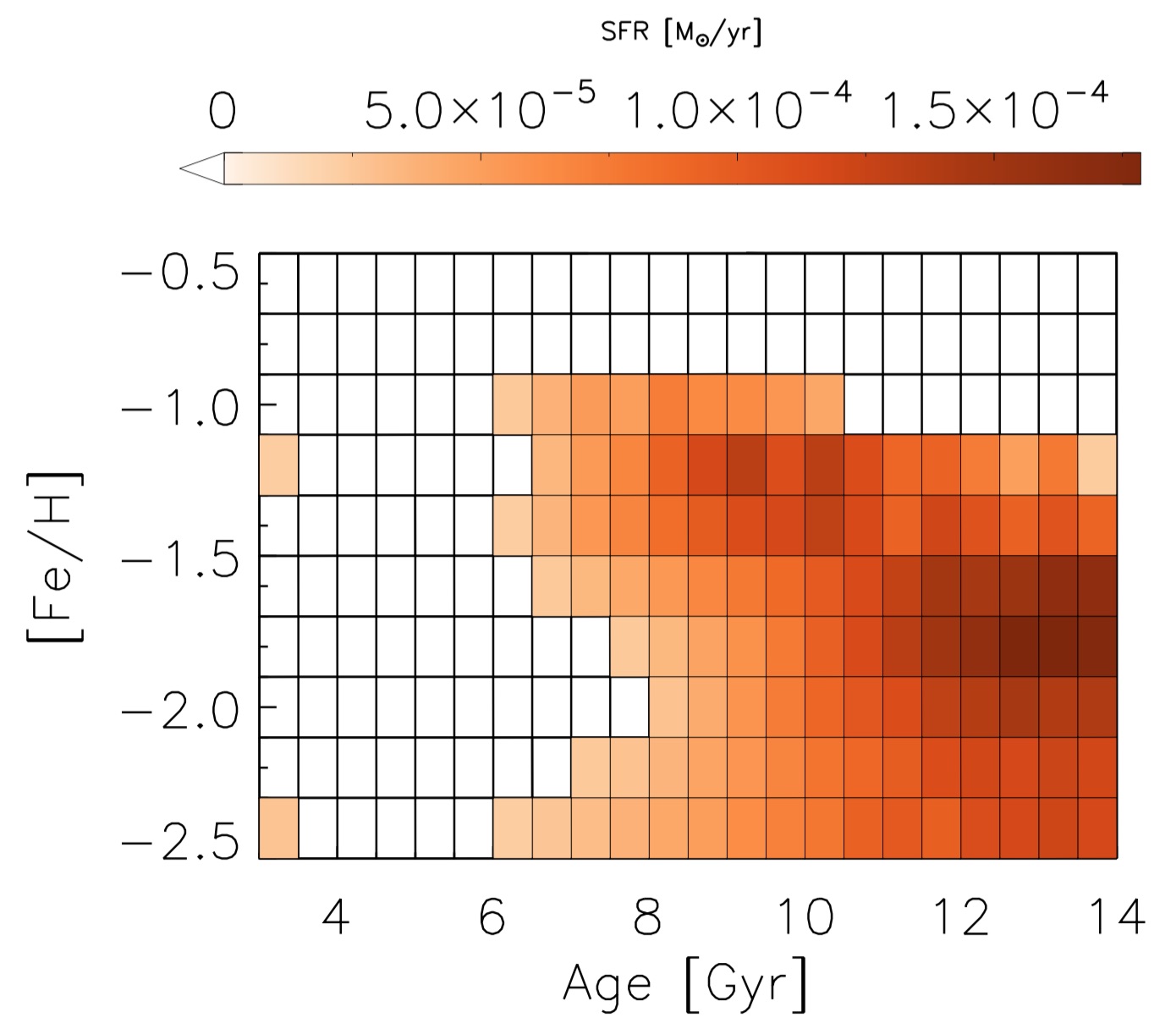}
    \caption{The age-metallicity distribution of star formation for Sculptor, recovered with {\small MORGOTH}, enhancing the HB contribution to the fit parameter.}
    \label{fig:3d-1}
\end{figure}

Our best fit CMD is shown in Fig.~\ref{fig:CMDscl}. The resemblance to the observations is, generally, satisfactory. There are, however, some differences in the details. We note that the MSTO and the subgiant branch are wider in our model than in the observed CMD, the model RGB has a slightly redder edge and there is a tail of blue HB stars in the model CMD that is not in the observed CMD. The origin of these problems is likely to be in the characterization of the faint CMD. Similar difficulties in reproducing the detailed structure of the main sequence region were also present in the original analysis of \citet{deBoer12}. The modelling of these faint stars is particularly sensitive to a reliable characterisation of observational conditions. As the available photometry reaches only one magnitude below the MSTO, an accurate estimate of all the uncertainties in this low signal-to-noise region can prove difficult. This problem highlights how crucial it is to have deep and well characterized dataset when modelling the resolved stellar populations in galaxies and how useful the information on the HB can be when such deep photometry is not available. In spite of these difficulties, and of the inclusion of additional constraints coming from the HB modelling, {\small MORGOTH} produced a best-fit model that is in line with what we know about the Sculptor dSph.

Looking at the best fit SFH (Fig.~\ref{fig:Scl}), we find that Sculptor's stellar population is predominantly old and metal poor, as found by \citet{deBoer12} and others before him. The star formation rates in the age-metallicity plane show a general evolution of the metal content with time, increasing by approximately one order  of magnitude over the SFH of the galaxy. Integrating our SFH in age and metallicity (adopting a Kroupa IMF) results in a total stellar mass, within an equivalent radius of 11 arcmin, of $5.32 \pm 0.53 \cdot 10^6 M_{\odot}$.


The detailed comparison with \citet{deBoer12} results is generally good, both in the age and metallicity distributions of the SFH. A further confirmation of the higher precision of our method is that, although our age bins are half the size of that used in \citet{deBoer12}, our uncertainties are noticeably smaller. Note that the uncertainties are expected to increase when reducing the parameter space bin size \citep{Dolphin02}.Our solution presents a more prominent tail of star formation at younger age. This is because \citet{deBoer12} masked the blue plume of stars above the MSTO, assuming them to be blue stragglers and thus unrelated to the SFH. Determining whether or not these stars come from a genuine recent star formation event or are blue stragglers is not easy and is beyond our analysis here.

A comparison with the previous RGB mass loss rate determination is also encouraging. The amount of mass loss inferred from our analysis is, overall, a few hundreds of solar mass higher than that measured by \citet{Salaris13}, but the two estimates are in good agreement within their uncertainties. The only significant discrepancy emerges at low metallicity, where we estimate a significantly higher mass loss. This high value regards stars with [Fe/H] $\lesssim -2.0$, which cause the blue HB tail in the model CMD. A possible explanation is that this difference comes from our imposition of a linear relation between integrated mass loss and metallicity. The estimate from \citet{Salaris13} shows a clear drop in the efficiency of mass loss for metal poor stars, greater than what would be expected by linearly extrapolating the values measured for more metal rich stars.

This discrepancy may indicate that our mass loss parametrisation is too simplistic. A higher order polynomial or a non-parametric representation \citep[similar to][]{Salaris13} could lead to a more accurate model. We stress, however, that increasing the number of degrees of freedom in our HB models is also likely to make the analysis more susceptible to degeneracies, and to lead to unrealistic solutions. Clearly, the optimal strategy to treat mass loss needs to be evaluated with care. The precise choice of mass loss parametrisation, however, is not critical to the conceptual validity of our method.

The modelling discrepancies stemming from the difficult observational characterisation of the faint CMD highlight a limitation of adding the HB to the CMD modelling. The Poissonian nature of the fit parameter we use causes the high number of MSTO stars to dominate the likelihood. Namely, when there is a difference between the SFH that best fits the MSTO and the one that best fits the HB then, without any intervention in the process, small improvements in the quality of the MSTO fit will be preferred to big improvements in the quality of the HB fit. A way to mitigate this issue would be to add an enhancement factor to the contribution of the HB to the final fit parameter, in a similar fashion with what is done with the RGB punishment region. We tested the effect of this approach by scaling the likelihood coming from the Hess bins of the HB region by the ratio of HB stars to non-HB stars. A drawback of this approach is that our fit parameter is not a rigorous Poissonian likelihood anymore. However, tests on synthetic populations revealed that this new fitting procedure improves the time resolution to 500 Myr. The SFH that is measured for Sculptor with this approach is showed in Fig.~\ref{fig:3d-1}. Although the overall shape of the SFH is in agreement with the SFH measured in Fig.~\ref{fig:SFHscl}, a closer look suggests the presence of two main subpopulations in Sculptor. The first one spans a range in age and metallicity of $11\,Gyr < t < 14\,Gyr$ and $-2.0<[Fe/H]<-1.5$. The second subpopulations has  $8\,Gyr < t <\,11 Gyr$ and $-1.5<[Fe/H]<-1.0$. The presence of these two components has already been noted using the spatial distribution of the HB morphology, the metallicity and velocity dispersion distributions on the RGB \citep{Tolstoy04,Zhu16}. The recovery of a bimodal SFH for Sculptor is supported by these previous detections and highlights the importance of having multiple independent constraints in the modelling of the CMD, expecially in the presence of systematics effects.

\section{Conclusions}
\label{conclusion}

We have presented and tested {\small MORGOTH}, to show it is capable of accurately determining the SFH of both simple and complex stellar populations by quantitatively taking into account all the major luminous features of the CMD, including the HB. It takes advantage of the internal structural similarities in evolved low mass stars, to model the helium burning phase of old populations in a flexible and computationally affordable manner. Simple tests with mock stellar populations reveal the benefit of having independent constraints on age and metallicity at the old ages, from the HB, increasing the time resolution of classical SFH determinations. Even with no constraints on the mass loss efficiency, our method is capable of a substantial improvement in the SFH precision, at the same as time measuring the total mass lost by RGB stars.

We also tested our method on observations of the Sculptor dSph. The SFH and mass loss estimates obtained are in good agreement with previous analyses \citep{deBoer12,Salaris13}, confirming the reliability of our approach. Our SFH measurements, including the HB, have smaller uncertainties compared with traditional, MSTO only, analysis techniques. Although a statistically consistent modelling of the CMD predicts a relatively simple SFH, enhancing the HB importance in the CMD fit reveals the two stellar subpopulations known to exist in Sculptor.

The detailed modelling of the HB of resolved stellar populations in galaxies opens interesting prospects for more distant surveys. Aside from the obvious advantage of having more accurate SFH measurements thanks to the additional age and metallicity indicators, we now also have the means to measure the amount of mass lost by RGB stars in external galaxies. Understanding RGB mass loss has been a stubborn long standing problem. In spite of decades of effort, a reliable characterization of this phenomenon has been difficult, even in apparently simpler systems like the Galactic globular clusters \citep[e.g.,][]{Catelan09}. The origin of this challenge lies in the nature of globular clusters, that are now known to contain a series of chemical peculiarities also reflected in their HB morphology \citep[e.g.,][]{Gratton11}. Dwarf galaxies,  on the other hand, seem to be free from these chemical anomalies \citep{Geisler07,Fabrizio15,Salaris13,savino15}, but the intrinsic spreads in the age and metallicity of their stellar populations has made the study of RGB mass loss equally difficult. Developing a method to study this phenomenon in complex stellar populations is, therefore, an important step towards a more complete understanding of both stellar and galactic evolution.

Additionally, a deeper knowledge of mass loss will allow us to obtain detailed SFHs, back to the earliest times, for a much larger number of galaxies.With current analysis techniques, accurate SFHs for the earliest stages of galaxy formation can only be measured if the faintest MSTOs are detected. This limits the maximum distance for this kind of study to the edge of the Local Group. If motivated assumptions can be made about the RGB mass loss, our method has the potential to measure the SFH from the HB alone, which is brighter than equivalent age MSTOs. Deep Hubble Space Telescope observations can already resolve the HB in galaxies outside the Local Group \citep[e.g.,][]{Dacosta10,Lianou13}. Next generation facilities, such as the Jame Webb Space Telescope, the European Extremely Large Telescope and the Thirty Meter Telescope, will be able to resolve HB stars for hundreds of galaxies within several Mpc from the Milky Way \citep{Brown08,Greggio12,Fiorentino17}, thus allowing accurate SFHs, back to the earliest times, for a large and diverse sample of resolved stellar systems, covering a range of environments, and over a cosmologically representative volume.

\section*{Acknowledgements}

We thank the anonymous referee who helped us improving this manuscript. A.S. wishes to thank I. Cabrera-Ziri and M. Sasdelli for the helpful discussions during a critical phase of the method's development; M. Cignoni and E. Sacchi for the helpful discussions on the modelling of Sculptor; A. Pietrinferni and S. Cassisi for providing us the theoretical tracks for high mass HB stars.




\bibliographystyle{mnras}
\bibliography{./Bibliography} 

\bsp	
\label{lastpage}
\end{document}